\documentclass[10pt,letterpaper]{article}
\pdfoutput=1

\usepackage{jcappub}
\usepackage{amsmath,amssymb,color}
\usepackage{enumerate,xstring}
\usepackage{jcappub}
\usepackage{amsmath}
\usepackage{amsfonts}
\usepackage{amssymb}
\usepackage{graphicx}
\usepackage{epsfig}
\usepackage{color}
\usepackage{multirow}
\usepackage{graphicx}
\usepackage{bigints}
\usepackage{todonotes,bm,etoolbox}
\usepackage{calligra}
\usepackage{hyperref}
\usepackage{tabularx}
\usepackage{booktabs}

\usepackage{tikz}

\def\ba#1\ea{\begin{align}#1\end{align}}
\def\bea{\begin{eqnarray}}
\def\eea{\end{eqnarray}}
\def\be{\begin{equation}}
\def\ee{\end{equation}}

\def\({\left(}
\def\){\right)}
\def\[{\left[}
\def\]{\right]}

\def\<{\left\langle}
\def\>{\right\rangle}

\def\comment#1{}

\def\eps{\epsilon}

\renewcommand{\v}[1]{\bm{#1}}

\def\vx{\v{x}}
\def\vk{\v{k}}

\def\highdm{{\rm High}\ \delta_m}
\def\lowdm{{\rm Low}\ \delta_m}
\def\highas{{\rm High}\ \A_s}
\def\lowas{{\rm Low}\ \A_s}
\def\highs{{\rm High}\ \sigma}
\def\lows{{\rm Low}\ \sigma}
\def\hodn{\langle{N}_g(M_h)\rangle}
\def\hodnlong{\langle{N}_g(M_h,z)\rangle}
\def\hodnloong{\langle{N}_g(M_h,z,\vx)\rangle}

\newcommand{\perm}[1]{ \expandafter\ifstrempty\expandafter{#1} {\mbox{perm.}} {\mbox{$#1$ perm.}} }

\def\A{\mathcal{A}}
\def\O{\mathcal{O}}

\newcommand{\fnl}{f_\textnormal{\textsc{nl}}}

\definecolor{RedWine}{rgb}{0.743,0,0}
\definecolor{RoyalBlue}{rgb}{0.25,.41,.88}
\definecolor{ForestGreen}{rgb}{.13,.54,.13}
\definecolor{Goldenrod}{rgb}{.85,.65,.13}

\newcommand{\bq}{\begin{eqnarray}}
\newcommand{\eq}{\end{eqnarray}}

\title{\huge Responses of Halo Occupation Distributions: \Large \newline a new ingredient in the halo model \& the impact on galaxy bias}

\author[a, d]{Rodrigo Voivodic}
\author[b,c,d]{and Alexandre Barreira}

\affiliation[a]{\small Departamento de F\'{\i}sica Matem\'atica, Instituto de F\'{\i}sica, Universidade de S\~ao Paulo,\\ R.  do  Mat\~ao  1371,  05508-090,  S\~ao Paulo, SP, Brazil}
\affiliation[b]{\small Excellence Cluster ORIGINS, Boltzmannstra\ss e 2, 85748 Garching, Germany}
\affiliation[c]{\small Ludwig-Maximilians-Universit\"at, Schellingstra\ss e 4, 80799 M\"unchen, Germany}
\affiliation[d]{\small Max-Planck-Institut f\"ur Astrophysik, Karl-Schwarzschild-Stra\ss e~1, 85748 Garching, Germany}

\emailAdd{rodrigo.voivodic@usp.br}
\emailAdd{alex.barreira@origins-cluster.de}

\date{\today}

\abstract{Halo occupation distribution (HOD) models describe the number of galaxies that reside in different haloes, and are widely used in galaxy-halo connection studies using the halo model (HM). Here, we introduce and study HOD {\it response functions} $R_\O^g$ that describe the {\it response} of the HODs to long-wavelength perturbations $\O$. The linear galaxy bias parameters $b_\O^g$ are a weighted version of $b_\O^h + R_\O^g$, where $b_\O^h$ is the halo bias, but the contribution from $R_\O^g$ is routinely ignored in the literature. We investigate the impact of this by measuring the $R_\O^g$ in separate universe simulations of the IllustrisTNG model for three types of perturbations: total matter perturbations, $\O = \delta_m$; baryon-CDM compensated isocurvature perturbations, $\O = \sigma$; and potential perturbations with local primordial non-Gaussianity, $\O \propto\fnl\phi$. Our main takeaway message is that the $R_\O^g$ are not negligible in general and their size should be estimated on a case-by-case basis. For stellar-mass selected galaxies, the responses $R_\phi^g$ and $R_\sigma^g$ are sizeable and cannot be neglected in HM calculations of the bias parameters $b_\phi^g$ and $b_\sigma^g$; this is relevant to constrain inflation using galaxies. On the other hand, we do not detect a strong impact of the HOD response $R_1^g$ on the linear galaxy bias $b_1^g$. These results can be explained by the impact that the perturbations $\O$ have on stellar-to-total-mass relations. We also look into the impact on the bias of the gas distribution and find similar conclusions. We show that a single extra parameter describing the overall amplitude of $R_\O^g$ recovers the measured $b_\O^g$ well, which indicates that $R_\O^g$ can be easily added to HM/HOD studies as a new ingredient.}

\begin{document}

\maketitle
\flushbottom

\section{Introduction}
\label{sec:introduction}

The {\it halo model} (HM) is one of the most popular frameworks to predict the statistics of the large-scale structure (LSS) of the Universe (see Ref.~\cite{cooray/sheth} for a review and Ref.~\cite{2020arXiv200914066M} for a recent online calculation tool). The core assumption of the HM is that all mass elements of a given tracer of the LSS, for example galaxies, are found in gravitationally-bound dark matter haloes. The description of the galaxy distribution then becomes effectively a two step process. The first step involves predicting the distribution of haloes in the Universe. Over the years, increasingly accurate $N$-body methods led to significant advances on this front: gravity-only codes agree to $\%$ levels nowadays \cite{2016JCAP...04..047S, 2015MNRAS.454.4208W}, and a number of accurate fitting formulae for dark matter halo profiles \cite{NFW, 1965TrAlm...5...87E}, abundance \cite{2002MNRAS.329...61S, 2008ApJ...688..709T} and bias \cite{2010ApJ...724..878T, lazeyras/etal} are also available. The second step involves describing the distribution of galaxies inside the haloes, which is an appreciably harder task given the many intertwined astrophysical processes that govern the formation of galaxies and their evolution.  One way to tackle this problem is via so-called {\it Halo Occupation Distribution} (HOD) models \cite{1997MNRAS.286..795K, 2000MNRAS.311..793B, 2000MNRAS.318..203S, 2000MNRAS.318.1144P, scoccimarro/etal:2001, 2002ApJ...575..587B, 2003ApJ...593....1B, 2004ApJ...609...35K, 2005ApJ...633..791Z, 2005ApJ...624..505Z, 2004MNRAS.353..189V} that describe the galaxy-halo connection via empirical parametrizations of aspects of the galaxy distribution inside halos such as their total number, radial profile and velocity profile.

One common application of the HM and HODs is the direct analytical calculation of $N$-point correlation functions of the galaxy distribution. In the simplest case of the 2-point correlation function, the calculation is split into the correlation between galaxies that live in the same halo and the correlation between galaxies in different haloes. The HM has also popularly been used to calculate $N$-point functions of the total matter distribution, although in this case, there is an important diffuse component that is not bound to haloes \cite{Voivodic3} (as well as other problems \cite{2016PhRvD..93f3512S}) that makes the calculation less accurate; see, however, Refs.~\cite{2015MNRAS.454.1958M, 2019MNRAS.488.2121C, 2020arXiv200901858M} for augmented versions of the HM (some including baryonic physics effects) that have been proving accurate enough for current parameter inference analyses. Another popular application of the HM and HODs is the fast generation of galaxy mock catalogues, in which galaxies are {\it painted} following some HOD formulae onto halo catalogues generated with gravity-only simulations. The fact that gravity-only simulations require far less computational resources than direct simulation of galaxy formation makes this a considerable computational advantage. The parameters of the HOD can then be iterated over to reproduce the statistics of some observed galaxy sample and learn about the galaxy-halo connection this way (see e.g.~Refs.~\cite{2002ApJ...577....1M, 2002MNRAS.329..246B, 2004ApJ...610...61Z, 2005ApJ...630....1Z, 2006ApJ...647..201C, 2008MNRAS.385.1257B, 2008MNRAS.387.1045W, 2009ApJ...707..554Z, 2011ApJ...736...59Z, 2011MNRAS.416.3033S, 2012ApJ...744..159L, 2013MNRAS.429.3604B, 2013MNRAS.428.2548K, 2014MNRAS.444..476R, 2015ApJ...806....2M, 2015MNRAS.454.1161Z, 2015MNRAS.449.1352C, 2015ApJ...807..152S, 2017MNRAS.467.3024L} for a number of examples of such analyses and Ref.~\cite{2018ARA&A..56..435W} for a review). 

The mean number of galaxies $\hodnlong$ that reside in haloes with mass $M_h$ at redshift $z$ is a central ingredient in HOD modelling. Haloes are normally assumed to contain zero or one central massive galaxy residing at the bottom of their potential wells. The haloes which are massive enough can further contain a number of lower mass satellite galaxies that orbit around the center (Fig.~\ref{fig:HOD_Mtot} below shows some the typical shapes of $\hodnlong$). For the most part in this paper we will consider galaxies as our specific example of a LSS tracer, but we note that the philosophy behind HOD modelling holds generically to other tracers as well. For example, in 21cm line intensity mapping, the relevant tracer is neutral hydrogen (HI), in which instead of a HOD number, one would use a function describing the total HI mass in haloes \cite{2017MNRAS.469.2323P, 2018ApJ...866..135V, 2019MNRAS.484.1007W}.

{\it In this paper, we focus on the dependence of HOD numbers on the large-scale environment.} The environmental dependence of HOD numbers has naturally been the subject of past works in the literature (see e.g.~Refs.~\cite{2003ApJ...593....1B, 2014PhDT.......126M, 2016arXiv160102693M, 2018ApJ...853...84Z, 2018MNRAS.480.3978A, 2019MNRAS.490.5693B, 2020MNRAS.493.5506H, 2020arXiv200804913H, 2020arXiv201105331H, 2020A&A...638A..60A, 2020arXiv201004182Y}), which are often carried out in the context of {\it assembly bias} (or secondary bias) \cite{gao/etal, 2007MNRAS.377L...5G, wechsler/etal, dalal/etal, 2014MNRAS.443.3044Z, 2016MNRAS.460.2552H, 2017JCAP...03..059L, 2018MNRAS.473.2486S, 2019MNRAS.484.1133C, 2020MNRAS.496.1182M, Tucci, Montero-Dorta2, 2020arXiv201008500J, 2020arXiv201004176S}, i.e., the dependence of galaxy clustering on properties beyond their host halo mass $M_{\rm h}$. In most of these studies, the environment is normally defined over length scales of a few times the typical virial radii of dark matter haloes ($R_{\rm env} \sim 1 - 10\ {\rm Mpc}/h$) on which the r.m.s.~fluctuations of the density field are of order unity. Here, we focus instead on the dependence of $\hodnlong$ on the environment defined over much larger linear scales ($R_{\rm env} \gtrsim 50 - 100\ {\rm Mpc}/h$, where the r.m.s.~fluctuations are much smaller than unity), which as we will see has important ramifications for how to interpret {\it galaxy biasing} using the HM. We will go into more details in sections below, but let us expose the context of our work briefly here.

At leading order in perturbation theory and for primordial adiabatic Gaussian scalar fluctuations, the local number density of galaxies can be written as (see Ref.~\cite{biasreview} for a comprehensive review)
\bq\label{eq:biasdef_intro}
n_g(z, \vx) = n_g(z)\Big[1 + b_{1}^g(z)\delta_m(z, \vx)\Big],
\eq
where $n_g(z, \vx)$ is the number density of galaxies in a region around position $\vx$ at redshift $z$, $n_g(z)$ is the global average over all positions and $\delta_m(z, \vx)$ is a linear total matter density perturbation.\footnote{We distinguish local from global averaged quantities by the presence or absence of the position $\vx$ in the argument. In this short exposition of the problem we also skip writing the contribution from {\it stochastic} terms that encapsulate the dependence of galaxy formation on the shorter-wavelength details of the environment.} In this equation, $b_{1}^g$ is called the linear local-in-matter-density galaxy bias parameter \cite{fry/gaztanaga:1983} and it quantifies how {\it responsive} the number density of galaxies is to the presence of the perturbation $\delta_m$. Using the HM, the global mean number of galaxies is given by integrating the global number density of haloes $n_h(M_h)$ weighted by the global HOD number $\hodn$ (we leave the dependence on $z$ implicit to simplify the notation):
\bq\label{eq:ng_hm_intro}
n_g = \int {\rm d}M_h n_h(M_h) {\hodn}.
\eq
From this equation, the local number of galaxies around position $\vx$ can be worked out by promoting both $n_h(M_h)$ and $\hodn$ to be {\it biased tracers} of $\delta_m$, i.e., to admit a description analogous to that in Eq.~(\ref{eq:biasdef_intro}): 
\bq\label{eq:nhHOD_hm_intro}
n_h(M_h,  \vx) &=& n_h(M_h)\big[1 + b_{1}^h(M_h) \delta_m(\vx)\big] \nonumber \\
\langle N_g(M_h, \vx)\rangle &=& \hodn\big[1 + R^g_{1}(M_h)\delta_m(\vx)\big],
\eq
where the parameters $b_{1}^h$ and $R^g_{1}$ are the equivalent of $b^g_{1}$ but for the dark matter halo abundance and the galaxy HOD number, respectively; we will refer to the parameter $R_1^g$ as the {\it HOD number response function}. It therefore follows that
\bq\label{eq:ng_hm_intro2}
n_g(\vx) = n_g \Bigg[1 + \frac{1}{n_g} \int {\rm d}M_h n_h(M_h)\hodn \Big(b_{1}^h(M_h) + R^g_{1}(M_h)\Big) \delta_m(\vx) \Bigg],
\eq
which can be contrasted with Eq.~(\ref{eq:biasdef_intro}) to derive the galaxy bias parameter as
\bq\label{eq:HM_bias}
b_{1}^g = \frac{1}{n_g} \int {\rm d}M_h n_h(M_h)\hodn \Big(b_{1}^h(M_h) + R^g_{1}(M_h)\Big).
\eq
The standard HM lore affirms that the bias of the galaxies is inherited directly from that of the host haloes weighted by the HOD number. This is recovered by the last equation only when $R^{g}_{1} = 0$, i.e., assuming that the mean galaxy HOD numbers $\hodn$ are the same at all positions in the Universe. There is however no prior reason to expect this to be the case; in fact, given that the dynamics of structure formation are sensitive to whether structures form in overdense, $\delta_m > 0$, or underdense regions, $\delta_m < 0$, it is actually physically plausible to expect the mean number of galaxies inside haloes to display some of this sensitivity.\footnote{This is the general finding of a number of recent studies in the literature \cite{2016arXiv160102693M, 2018ApJ...853...84Z, 2018MNRAS.480.3978A, 2019MNRAS.490.5693B, 2020MNRAS.493.5506H, 2020arXiv200804913H, 2020arXiv201105331H} using definitions of the environment on $1 - 10\ {\rm Mpc}/h$ scales. For example, in very dense environments, mergers take place more frequently, which boosts the number of substructures at fixed halo mass. In contrast, in less dense regions \cite{2020A&A...638A..60A}, a larger fraction of the build up of halo mass is due to the infall of surrounding diffuse matter. In this paper, we focus on environments defined over large linear scales ($\gtrsim 50 - 100\ {\rm Mpc}/h$, where $\delta_m \ll 1$) relevant to the calculation of $b_1^g$ using the HM.} This is an issue that is briefly alluded to in Refs.~\cite{2004PhRvD..69h3524A, biasreview}, but which to the best of our knowledge has never been specifically addressed and tested in practice before. {\it This motivates our main goal in this paper}, which is to measure the size and shape of HOD response functions like $R^g_1$ using galaxy formation simulations and discuss the corresponding implications to the modelling of large-scale structure using the HM and HODs.

Concretely, in this paper we measure responses of HOD numbers $\hodn$ using separate universe simulations of the IllustrisTNG galaxy formation model \cite{Pillepich:2017jle, 2017MNRAS.465.3291W, Nelson:2018uso}. Separate universe simulation is a technique that allows to efficiently measure the responses of LSS quantities to long-wavelength perturbations by absorbing the effects of the perturbations into appropriate modifications to the cosmology of the simulation and its box size. These simulations have been run in previous works for the case of total matter density perturbations $\delta_m$ \cite{2019MNRAS.488.2079B}, primordial baryon-CDM compensated isocurvature perturbations (CIP) $\sigma$ \cite{2020JCAP...02..005B} and potential perturbations with local primordial non-Gaussianity (PNG) $\fnl\phi$ \cite{2020arXiv200609368B} (we describe these perturbations and simulations with more detail in the next section). We measure the responses of the HOD numbers $\hodn$ of the simulated galaxies to these three types of perturbations, as well as the response of the gas mass inside haloes for a few example elements tracked in IllustrisTNG.  Our main finding is that the HOD response functions are not negligible in general and can lead to sizeable effects on the corresponding galaxy bias parameters calculated using the HM. We also see that modelling the HOD responses as low order polynomials ($n \leq 2$) in $\log M_{h}$ recovers the correct value of the bias parameters. This all motivates the incorporation of HOD responses as an extra ingredient in traditional HOD studies. 

The rest of this paper is organized as follows. In Sec.~\ref{sec:sepuni}, we introduce the three types of long-wavelength perturbations we consider, as well as the separate universe simulations that we use to measure the corresponding HOD number responses. Section \ref{sec:results} contains our main results on the measurements of $\hodn$ and its response functions, and the illustration of the general necessity to take the responses into account within the HM to correctly recover the measured bias parameters. We show results for samples with varying total and stellar mass cuts, as well as for the distribution of hydrogen, carbon and oxygen as a few example gas elements tracked in IllustrisTNG. We summarise and conclude in Sec.~\ref{sec:summary}.  
     
\section{Responses and Separate Universe simulations}
\label{sec:sepuni}

Given any quantity $Q$ in LSS, its response functions $R^{Q}_{\O}$ can be defined with all generality via the expansion
\bq\label{eq:response_def}
Q(z, \vx) = Q(z) \Bigg[1 + \sum_\O R^{Q}_{\O}(z) \O(z, \vx)\Bigg] + \eps^{Q}(z, \vx),
\eq
where $Q(z, \vx)$ denotes the value of the quantity measured in some local volume around $\vx$, $Q(z)$ is the corresponding cosmic average and the sum runs over all long-wavelength perturbations $\O$ that can influence the quantity's local value. This equation makes apparent the physical meaning of the $R^{Q}_{\O}$ as the {\it response} (hence the name) of $Q$ to the presence of the perturbations $\O$. The response functions depend generically on redshift, as well as any other variable $Q$ may also depend on. In this equation, the perturbations $\O$ are assumed to be sufficiently long-wavelength that they can be treated perturbatively \cite{Bernardeau/etal:2002}. In Eq.~(\ref{eq:response_def}), $\eps^{Q}(z, \vx)$ is added to absorb all of the dependence of the value of $Q(z, \vx)$ on the shorter-wavelength part of the perturbations $\O$. We note that although this so-called stochastic contribution $\eps^{Q}(z, \vx)$ does not correlate with any of the $\O$, it is still important to take it into account in practical descriptions of the statistics of $Q(z, \vx)$; in this paper we focus solely on the response functions inside the squared bracket. We refer the interested reader to Ref.~\cite{biasreview} for a comprehensive review of the application of the expansion of Eq.~(\ref{eq:response_def}) to galaxy number densities as the quantity $Q$ (popularly known as the galaxy bias expansion) and to Refs.~\cite{response, 2014JCAP...05..048C, 2015JCAP...09..028C, akitsu/takada/li, responses1, responses2, 2018JCAP...02..022L} for an application of the same ideas to matter/galaxy power spectra. 

In this paper we look into the responses to three types of long-wavelength perturbations $\O$:
\begin{enumerate}

\item The first are total matter density perturbations $\O = \delta_m(z, \vx)$, which originate from adiabatic energy density fluctuations in the early Universe;

\item The second are baryon-CDM compensated isocurvature perturbations (CIP) $\O = \sigma(\vx)$ \cite{2003PhRvD..67l3513G, 2009PhRvD..80f3535G, 2010ApJ...716..907H, grin/dore/kamionkowski, 2014PhRvD..89b3006G, 2016PhRvD..94d3534H, 2016PhRvD..93d3008M, 2017PhRvD..96h3508S, 2019MNRAS.485.1248S, 2019PhRvD.100j3528H, 2019PhRvD.100f3503H, 2020JCAP...07..049B}, which are characterized by perturbations in the baryon density that are exactly compensated by the cold dark matter (CDM) to leave the total matter distribution unchanged. The power spectrum of CIPs, which cannot be generated by single-field inflation models, is remarkably poorly constrained by the CMB data \cite{2020A&A...641A..10P}. This has motivated a number of studies of their impact on galaxy statistics, which have shown to be possible to significantly improve on the current observational bounds \cite{2019MNRAS.485.1248S, 2019PhRvD.100f3503H, 2019PhRvD.100j3528H, 2020JCAP...07..049B}.

\item The third type are perturbations of the primordial gravitational potential in local PNG cosmologies, $\O = \fnl\phi(\vx)$; the value of $\fnl$ describes the amount of non-Gaussianity via $\phi = \phi_{\rm G} + \fnl\left[\phi_{\rm G}^2 - \langle\phi_{\rm G}^2\rangle\right]$, where $\phi_{\rm G}$ is a Gaussian random variable and $\langle\rangle$ denotes ensemble average \cite{2001PhRvD..63f3002K}. The tightest observational bounds constrain $\fnl =0.9 \pm 5.1$ \cite{2020A&A...641A...9P} and there is significant interest in improving further on this bound given the power of a detection of $\fnl \neq 0$ to rule out multi-field models of inflation \cite{maldacena:2003, 2004JCAP...10..006C, 2011JCAP...11..038C, Tanaka:2011aj, baldauf/etal:2011, conformalfermi, CFCpaper2, 2015JCAP...10..024D};
\end{enumerate}

We will focus on the responses of galaxy number densities $n_g$ and galaxy HOD numbers $\hodn$. For these two quantities and for the three types of perturbations we consider, the corresponding response expansions are given, respectively, by (recall, we distinguish the local from the global quantities by the presence of the position $\vx$ in the arguments)
\bq\label{eq:response_exp_ng} 
n_g(\vx, z) &=& n_g(z) \Big[1 + b_{1}^g(z)\delta_m(z, \vx) + b_{\sigma}^g(z)\sigma(\vx) + b_{\phi}^g(z)\fnl\phi(\vx)\Big]
\eq
and
\bq\label{eq:response_exp_Ng}
\hodnloong &=& \hodnlong \Big[1 + R^{g}_{1}(M_h, z)\delta_m(z, \vx) + R^{g}_{\sigma}(M_h, z)\sigma(\vx) + R^{g}_{\phi}(M_h, z)\fnl\phi(\vx)\Big], \nonumber \\
\eq
where we skipped writing the stochastic contributions explicitly since we do not study them here. 

In the remainder of this section, we outline the main specifications of the separate universe simulations that we use in this paper to measure $b_\O^g$ and $R_\O^g$. The separate universe simulation technique builds on top of the assumption that the physics that determine the quantity $Q$ act on sufficiently small scales that they regard the large-scale perturbations $\O$ as changes to the background (this is also called the peak-background split approach \cite{kaiser:1984, bardeen/etal:1986}).  Specifically, the separate universe ansatz states that {\it the formation of structures locally inside long-wavelength perturbations is equivalent to the formation of structures globally in an appropriately modified cosmology.} For the case of $\O = \delta_m$ the modified cosmology has a different mean background total matter density, for $\O = \sigma$ the cosmology has different relative amplitudes of the cosmic fractions of baryons and CDM, and for $\O = \fnl\phi$ the modified cosmology has a different amplitude $\A_s$ of the primordial scalar power spectrum. 

\begin{figure}[t!]
	\centering
	\includegraphics[width=\textwidth]{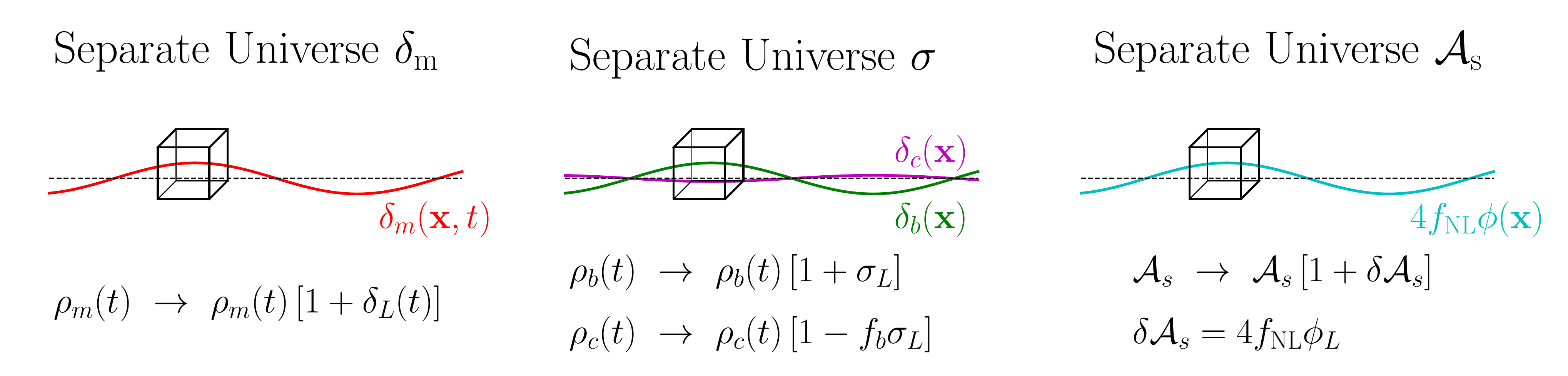}
	\caption{Summary sketch of the separate universe approach to the three types of long-wavelength perturbations considered in this paper. On the left, local structure formation inside long-wavelength matter perturbations $\delta_m$ is equivalent to global structure formation in a separate cosmology with modified background matter density ${\rho}_m$. In the center, local structure formation inside long-wavelength CIP perturbations $\sigma$ is equivalent to global structure formation with modified background baryon $\rho_b$ and CDM $\rho_c$ densities, at fixed total matter density. On the right, local structure formation inside long-wavelength primordial gravitational potential perturbations with local PNG is equivalent to global structure formation in a cosmology with modified amplitude of the primordial scalar perturbation power spectrum $\mathcal{A}_s$.} 
\label{fig:sepuni}
\end{figure}

The simulations we use here have been described in previous works, and so we shall be brief in descriptions and refer the interested reader to the cited literature for more details and derivations. Next, we describe first the main numerical aspects that are common to all simulations and then comment in turn on the specifics of the three types of separate universe simulations we consider; the main idea and numerical details are summarized in Fig.~\ref{fig:sepuni} and Table \ref{tab:params}. At the end of this section, we show the measurements of the halo bias $b_\O^h$ from the separate universe simulations (the equivalent of the $b_\O^g$ in Eq.~(\ref{eq:response_exp_ng}), but for halo number counts).

\subsection{Numerical details}\label{sec:details}

The simulations we use in this work were run with the moving-mesh code {\sc AREPO} \citep{2010MNRAS.401..791S, 2016MNRAS.455.1134P} and the IllustrisTNG galaxy formation model \citep{2017MNRAS.465.3291W, Pillepich:2017jle,Nelson:2018uso}. The latter is an improved version of the Illustris model \citep{2014MNRAS.445..175G, 2014MNRAS.444.1518V} and it includes prescriptions for gas cooling and ionization, star formation and feedback, and black hole growth and feedback; Refs.~\cite{2018MNRAS.480.5113M, Pillepich:2017fcc, 2018MNRAS.477.1206N, 2018MNRAS.475..676S, Nelson:2017cxy} present and discuss a number of the first key results obtained with IllustrisTNG.

In our main results, we consider simulations run at two mass resolutions. One, which we label as TNG100-1.5, corresponds to a box size $L_{\rm box} = 75{\rm Mpc}/h$ and $N_p = 2\times 1250^3$ tracer elements; and another, labeled TNG300-2, with $L_{\rm box} = 205{\rm Mpc}/h$ and $N_p = 2\times 1250^3$. The initial conditions were generated at $z = 127$ with the {\sc N-GenIC} code \citep{2015ascl.soft02003S} using the Zel'dovich approximation; the input linear matter power spectrum was calculated with the {\sc CAMB} code \citep{camb}. For the most part in this paper, we show results from the full hydrodynamical simulations with IllustrisTNG (dubbed Hydro), although in Sec.~\ref{sec:results_totmass} we display also results from their gravity-only counterparts (with half the number of mass elements; we dub these as Gravity). There, we consider also an additional set of gravity-only separate universe simulations available for the cases $\O = \delta_m$ and $\O = \fnl\phi$, with $L_{\rm box} = 560{\rm Mpc}/h$ and $N_p = 1250^3$; we label this resolution simply as $L_{\rm box}  \approx 800 {\rm Mpc}$.

Gravitationally bound haloes are identified with a Friends-of-Friends (FOF) code run on the dark matter tracer particles with a linking length $b = 0.2$ times the mean interparticle distance. Inside each halo, subhaloes are found using the {\sc SUBFIND} algorithm \cite{2001MNRAS.328..726S}. As is standard, we define the {\it main central} subhalo to be that which resides at the bottom of the potential of the FOF object, with all of the remaining subhaloes being called {\it satellites}. When quoting mass values for these structures we always consider the summed mass from all elements that belong to the halo/subhalo. For example, the total stellar mass of a halo is the summed mass of all star particles that are assigned to the halo, which includes those that also belong to its subhalos. Following the standard nomenclature in IllustrisTNG-related literature, we refer to subhalos that contain any mass in stars as {\it galaxies}.

\begin{table}
\centering
\begin{tabular}{@{}lccccccccccc}
\hline\hline
\rule{0pt}{1\normalbaselineskip}
Name &\ \ $\Omega_{m0}$ & \ \ $\Omega_{b0}$ & \ \ $\Omega_{c0}$ & \ \ $\Omega_{\Lambda0}$ & \ \ $h$ & \ \ $n_s$ & \ \ $\mathcal{A}_s$
\\
\hline
\rule{0pt}{1\normalbaselineskip}
${\rm Fiducial}$ &\ \ $0.3089$ & \ \ $0.0486$ & \ \ $0.2603$ & \ \ $0.6911$ & \ \ $0.6774$ & \ \ $0.967$ & \ \ $2.068 \times 10^{-9}$ 
\\
\\
\,\,$\highdm$ &\ \ $0.3194$ & \ \ $0.0502$ & \ \ $0.2692$ & \ \ $0.7146$ & \ \ $0.6662$ & \ \ {\footnotesize ${\rm Fiducial}$} & \ \ {\footnotesize ${\rm Fiducial}$}
\\
\,\,$\lowdm$ &\ \ $0.2991$ & \ \ $0.0471$ & \ \ $0.2520$ & \ \ $0.6691$ & \ \ $0.6884$ & \ \ {\footnotesize ${\rm Fiducial}$} & \ \ {\footnotesize ${\rm Fiducial}$}
\\
\\
\,\,${\highs}$ &\ \ {\footnotesize ${\rm Fiducial}$} & \ \ $0.0510$ & \ \ $0.2579$ & \ \ {\footnotesize ${\rm Fiducial}$} & \ \ {\footnotesize ${\rm Fiducial}$} & \ \ {\footnotesize ${\rm Fiducial}$} & \ \ {\footnotesize ${\rm Fiducial}$}
\\
\,\,${\lows}$ &\ \ {\footnotesize ${\rm Fiducial}$} & \ \ $0.0462$ & \ \ $0.2627$ & \ \ {\footnotesize ${\rm Fiducial}$} & \ \ {\footnotesize ${\rm Fiducial}$} & \ \ {\footnotesize ${\rm Fiducial}$} &  \ \ {\footnotesize ${\rm Fiducial}$}
\\
\\
\,\,${\highas}$ &\ \ {\footnotesize ${\rm Fiducial}$} & \ \ {\footnotesize ${\rm Fiducial}$} & \ \ {\footnotesize ${\rm Fiducial}$} & \ \ {\footnotesize ${\rm Fiducial}$} & \ \ {\footnotesize ${\rm Fiducial}$} & \ \ {\footnotesize ${\rm Fiducial}$} & \ \ $2.171 \times 10^{-9}$
\\
\,\,${\lowas}$ &\ \ {\footnotesize ${\rm Fiducial}$} & \ \ {\footnotesize ${\rm Fiducial}$} & \ \ {\footnotesize ${\rm Fiducial}$} & \ \ {\footnotesize ${\rm Fiducial}$} & \ \ {\footnotesize ${\rm Fiducial}$} & \ \ {\footnotesize ${\rm Fiducial}$} & \ \ $1.965 \times 10^{-9}$
\\
\hline
\hline
\end{tabular}
\caption{Cosmological parameters of the simulations used in this paper. The $\highdm$ and $\lowdm$ simulations are used to measure responses to $\delta_m$; $\highs$ and $\lows$ are used to measure the responses to CIP perturbations $\sigma$; and $\highas$ and $\lowas$ are used to measure the responses to $\delta\A_s = 4\fnl\phi$ (cf.~Secs.~\ref{sec:sepunidm}, \ref{sec:sepunisigma} and \ref{sec:sepunias}). These cosmologies were simulated with {\sc Arepo} and IllustrisTNG at two numerical resolutions: TNG100-1.5 with $L_{\rm box} = 75\ {\rm Mpc/h}$ and TNG300-2 with $L_{\rm box} = 205\ {\rm Mpc/h}$, both with $N_p = 2\times 1250^3$ mass elements. For all cases except ${\highs}$ and ${\lows}$, there is an additional numerical resolution for gravity-only dynamics with $L_{\rm box} = 560\ {\rm Mpc/h}$ and $N_p = 1250^3$, which we label simply as $L_{\rm box} \approx 800 {\rm Mpc}$. In all cosmologies, we approximate neutrinos as massless (see Refs.~\cite{2018PhRvD..97l3526C, Chiang:2016vxa, Jamieson:2018biz} for separate universe simulations where neutrinos and other scale-dependent effects arising in quintessence cosmologies are taken into account).}
\label{tab:params}
\end{table}

\subsection{The $\delta_m$ Separate Universe}\label{sec:sepunidm}

According to the separate universe ansatz, long-wavelength total matter perturbations are regarded by smaller-scale structure formation as a uniform change in the total matter density. Hence, the measurements of the responses to $\delta_m$ can be carried out by comparing the results from the Fiducial cosmology with those from cosmologies with modified cosmic total matter density. If $\delta_L(t)$ denotes the amplitude of the total (linear) matter perturbation at physical time $t$, then the separate universe cosmology is characterized by
\bq\label{eq:sepunidm}
\tilde{\rho}_m(t) = \rho_m(t)\left[1 + \delta_L(t)\right],
\eq
where a tilde indicates a quantity in the separate universe cosmology. The corresponding cosmological parameters are listed in Tab.~\ref{tab:params}; the cosmologies $\highdm$ and $\lowdm$ mimic the effects of positive and negative amplitude perturbations with present-day values $\delta_L^{\rm High}(z=0) = +0.05$ and $\delta_L^{\rm Low}(z=0) = -0.05$, respectively (note that $\delta_L$ evolves with time according to linear theory). The size of the amplitude of these perturbations is chosen as a compromise between having it to be large enough for the measurements to be high signal-to-noise, but small enough to keep higher-order terms (e.g.~$\O = \delta_m^2$) in the response expansion negligible. 

The different values of $h$ and the different relation between redshift and physical time in the Fiducial, $\highdm$ and $\lowdm$ cosmologies imply a number of important adjustments to the box size, FoF linking length, output times and setting up of the initial conditions. These have been discussed at length in previous works \cite{li/hu/takada, 2014PhRvD..90j3530L, wagner/etal:2014, CFCpaper2, baldauf/etal:2015, response, lazeyras/etal, li/hu/takada:2016, 2018PhRvD..97l3526C, 2019MNRAS.488.2079B, 2020arXiv200609368B}, to which we refer the reader for more details (see e.g.~Refs.~\cite{2019MNRAS.488.2079B, 2020arXiv200609368B} for the descriptions of the actual separate universe simulations we use here). When we quote units with $h$ factors, we always convert to the Fiducial cosmology.

From Eqs.~(\ref{eq:response_exp_ng}) and (\ref{eq:response_exp_Ng}), the bias parameter $b_{1}^g$ and the $\hodn$ response $R_{1}^{g}$ can be formally defined as
\bq\label{eq:b1R1defs}
b_{1}^{g}(z) = \frac{{\rm dln}\ n_g(z)}{{\rm d}\delta_L(z)} \bigg|_{\delta_{L}(z)=0} \ \ \ \ ; \ \ \ \ R_{1}^{g}(M_h,z) = \frac{{\rm dln}\ \hodnlong}{{\rm d}\delta_L(z)} \bigg|_{\delta_{L}(z)=0},
\eq
which we evaluate by finite differencing the results from the Fiducial, $\highdm$ and $\lowdm$ simulations. Concretely, we evaluate $R_{1}^{g}(M_h,z)$ as
\bq\label{eq:R1fd}
R_{1}^{g}(M_h,z) = \frac{R_{1}^{g, \highdm}(M_h,z) + R_{1}^{g, \lowdm}(M_h,z)}{2},
\eq
with
\bq\label{eq:R1fd_2}
R_{1}^{g, \highdm}(M_h,z) &=& \frac{1}{\delta_L^{\rm High}(z)} \left[\frac{\hodnlong^{\rm \highdm}}{\hodnlong^{\rm Fiducial}} - 1\right] ,  \nonumber \\
R_{1}^{g, \lowdm}(M_h,z) &=& \frac{1}{\delta_L^{\rm Low}(z)} \left[\frac{\hodnlong^{\rm \lowdm}}{\hodnlong^{\rm Fiducial}} - 1\right] ,
\eq
where $\hodn^{\rm Cosmology}$ denotes the galaxy HOD number measured in the corresponding cosmology. The calculation of $b_1^g$ is done analogously by finite differencing the total number of galaxies in the simulation box, instead of the number of galaxies inside haloes with mass $M_h$. In doing so, one actually measures the so-called Lagrangian value $b_1^{g, {\rm Lag.}}$, which is related to the (Eulerian) value that we consider in this paper by $b_1^g = b_1^{g, {\rm Lag.}} + 1$.\footnote{This follows from the relation between number densities in Eulerian and Lagrangian space.} 

An important point to note is that we have only one realization of the initial conditions, which prevents us from quoting error bars on our measurements in a statistical ensemble sense. The size of the simulation boxes also makes it hard to estimate errors via resampling of subvolumes. The responses $\highdm$ and $\lowdm$ are the forward and backward first-derivatives and are therefore the same in theory. In practice, however, numerical noise and binning effects can drive some differences, and so we take the different between $\highdm$ and $\lowdm$ as a rough estimate of the error in our measurements  (Ref.~\cite{2020arXiv200609368B} verified that, at least for $b_1^g$, these error bars are comparable in size to the estimate based on jackknife resampling presented in Ref.~\cite{baldauf/etal:2015}). 

\subsection{The $\sigma$ Separate Universe}\label{sec:sepunisigma}

A long-wavelength CIP is regarded by the structures within it as a uniform change to the cosmic fraction of baryons $\Omega_b$ and CDM $\Omega_c$, at fixed total matter $\Omega_m$. With the convention that $\sigma > 0$ corresponds to more baryons (less CDM), then the responses to CIPs can be evaluated using separate universe cosmologies characterized by
\bq\label{eq:sepunisigma}
\tilde{\Omega}_{b0} = \Omega_{b0}\left[1 + \sigma_L\right] \ \ \ {;}\ \ \ \tilde{\Omega}_{c0} = \Omega_{c0}\left[1 - f_b \sigma_L\right],
\eq
where $\sigma_L$ is the amplitude of the long-wavelength CIP and $f_b$ is the ratio of baryon-to-CDM density in the fiducial cosmology; note that $\sigma_L$ is constant in time because on large-scales the only relevant force is gravity, which acts equally on baryons and CDM. The implementation of these separate universe simulations is straightforward with the only difference relative to the Fiducial being just that the initial linear matter power spectrum should be generated for different values of $\Omega_{b0}$ and $\Omega_{c0}$ (see also Ref.~\cite{2020arXiv201101037K} for a numerical study of CIPs in which different transfer functions are used to generate the initial distribution of baryons and CDM). We consider two CIP cosmologies, $\highs$ and $\lows$, which correspond to $\sigma_L = +0.05$ and $\sigma_L = -0.05$, respectively (cf.~Tab.~\ref{tab:params}); these are the same simulations used previously in Ref.~\cite{2020JCAP...02..005B}.

The response functions are defined as
\bq\label{eq:bsigmaRsigmadefs}
b_{\sigma}^{g}(z) = \frac{{\rm dln}\ n_g(z)}{{\rm d}\sigma_L} \bigg|_{\sigma_L=0} \ \ \ \ ; \ \ \ \ R_{\sigma}^{g}(M_h,z) = \frac{{\rm dln}\ \hodnlong}{{\rm d}\sigma} \bigg|_{\sigma_L=0},
\eq
and we evaluate them via finite-differencing analogously to $b_1^g$ and $R_1^{g}$ above.

\subsection{The $\A_s$ Separate Universe}\label{sec:sepunias}

In cosmologies with local PNG ($\fnl \neq 0$), long-wavelength perturbations of the primordial gravitational potential are regarded by local structure formation inside them as a rescaling of the amplitude $\A_s$ of the scalar primordial power spectrum \cite{dalal/etal:2008, slosar/etal:2008}. Concretely, using the separate universe picture, it can be shown that the responses to these types of perturbations can be evaluated by comparing the Fiducial cosmology with cosmologies characterized by\footnote
{
Local PNG generates a primordial bispectrum that peaks in the squeezed limit, i.e., it generates a coupling between long-wavelength and short-wavelength modes. Concretely, the long-wavelength mode acts as a modified \emph{background} to the local, short-scale power spectrum, which gets modulated as (see Refs.~\cite{dalal/etal:2008, slosar/etal:2008} or Sec.~7 of Ref.~\cite{biasreview} for the squeezed bispectrum derivation):
\bq
P_{\phi\phi}(k_{S}, \vx) = \big[1 + 4\fnl\phi(k_{L})e^{i\vx.\vk_{L}}\big]P_{\phi\phi}(k_{S}),
\eq
where $P_{\phi\phi}(k_{S}, \vx)$ is the local power spectrum of the primoridal gravitational potential,  $P_{\phi\phi}(k_{S})$ is its spatial average, and $k_{S}$ and $k_{L}$ denote short- and long-wavelength modes, respectively. Thus, on distance scales sufficiently smaller than $1/k_{L}$, the effects on structure formation are equivalent to a rescaling of the amplitude of the primordial scalar power spectrum by $1 + \delta\A_s$, where $\delta\A_s = 4\fnl\phi_L$.
}
\bq
\tilde{\A}_s = {\A}_s \left[1 + \delta\A_s\right]\ \ , \ \ {\rm where}\ \ \delta\A_s = 4\fnl\phi_L,
\eq
with $\phi_L$ being the amplitude of the primordial potential perturbation. The amplitude $\A_s$ is the only parameter that differs relative to the fiducial cosmology, which makes these separate universe simulations also straightforward to setup: the only change is at the level of the initial conditions, which should be generated with an initial matter power spectrum rescaled by $1 + \delta\A_s$. The two cosmologies we consider, $\highas$ and $\lowas$, are characterized by $\delta\A_s = +0.05$ and $\delta\A_s = -0.05$, respectively (cf.~Tab.~\ref{tab:params}); these are the same simulations used previously in Ref.~\cite{2020arXiv200609368B}.

Similarly to the other two cases above, the responses to $\mathcal{A}_s$ are defined as
\bq\label{eq:bphiRphidefs}
b_{\phi}^{g}(z) = 4\frac{{\rm dln}\ n_g(z)}{{\rm d}\delta\A_s} \bigg|_{\delta\A_s=0} \ \ \ \ ; \ \ \ \ R_{\phi}^{g}(M_h,z) = 4 \frac{{\rm dln}\ \hodnlong}{{\rm d}\delta\A_s} \bigg|_{\delta\A_s=0},
\eq
and we evaluate them via finite-differencing analogously to the responses to $\delta_m$ and $\sigma$ described already above. The factor of $4$ in Eqs.~(\ref{eq:bphiRphidefs}) accounts simply for the fact that the responses multiply $\fnl\phi$ in Eqs.~(\ref{eq:response_exp_ng}) and (\ref{eq:response_exp_Ng}), but the amplitude rescaling is given by $\delta\A_s = 4\fnl\phi$.

\subsection{The halo bias parameters}

\begin{figure}
    \centering
    \includegraphics[width=\textwidth]{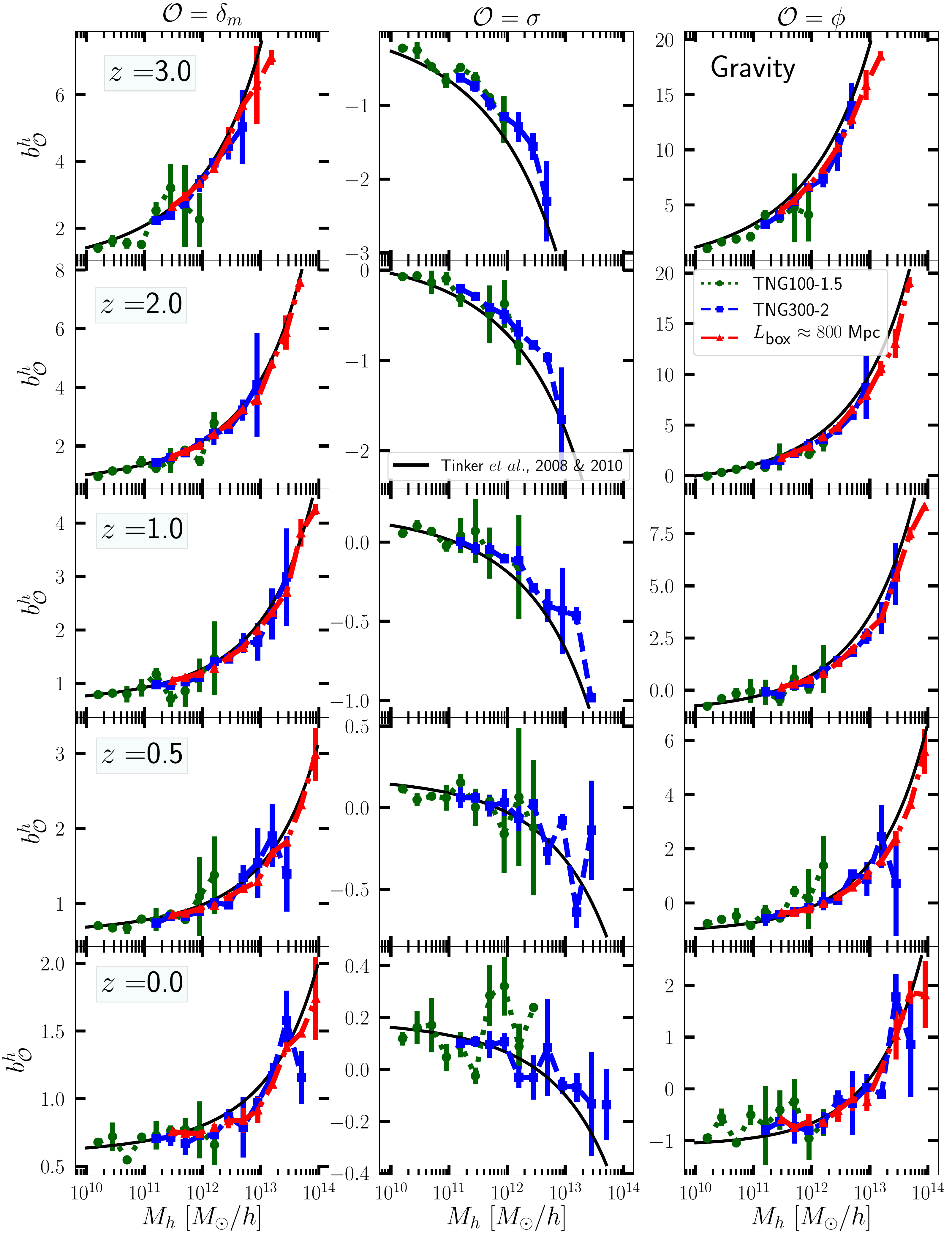}
    \caption{Halo bias parameters $b_1^h$ (left), $b_\sigma^h$ (center) and $b_\phi^h$ (right) as a function of total halo mass at different redshifts (different panels) and for the three gravity-only resolutions we consider in this paper, as labeled. The bias parameters are defined by the analog of Eq.~(\ref{eq:response_exp_ng}) for the halo (not galaxy) number density. In the left panels, the solid black line shows the prediction from the $b_1^h$ fitting formula of Ref.~\cite{2010ApJ...724..878T}. In the center and right panels, the solid line shows the result obtained by applying the separate universe ansatz to the halo mass function fitting formula of Ref.~\cite{2008ApJ...688..709T}.}
    \label{fig:bias_halos}
\end{figure}

Figure \ref{fig:bias_halos} shows the halo mass and redshift dependence of the linear halo bias parameters $b_1^h$, $b_\sigma^h$ and $b_\phi^h$. These are defined and measured analogously to the galaxy bias parameters, but considering halo instead of galaxy number densities. The result is shown for the three numerical resolutions available for gravity-only dynamics, which all agree to within the precision of our measurements. In the left panels for $b_1^h$, the black solid line shows the prediction from the fitting formula of Ref.~\cite{2010ApJ...724..878T}, whereas in the center and right panels, for $b_\sigma^h$ and $b_\phi^h$, respectively, the solid black line shows the result obtained using the halo mass function fitting formula of Ref.~\cite{2008ApJ...688..709T} with the definition of $b_\sigma^g$ and $b_\phi^g$ in Eqs.~(\ref{eq:bsigmaRsigmadefs}) and (\ref{eq:bphiRphidefs}), respectively. Specificaly, we use the formula of Ref.~\cite{2008ApJ...688..709T} to evaluate the mass function at the corresponding cosmologies in Tab.~\ref{tab:params}, and then compute the derivatives in Eqs.~(\ref{eq:bsigmaRsigmadefs}) and (\ref{eq:bphiRphidefs}) using finite differences.

These measurements have been presented and discussed with more detail in Ref.~\cite{2020arXiv200609368B} for $b_1^h$ and $b_\phi^h$ and Ref.~\cite{2020JCAP...02..005B} for $b_\sigma^h$ (see also Ref.~\cite{2020arXiv201101037K}), to which we refer the reader for more details. We display them here for completeness and because the halo bias parameters are a crucial ingredient in the HM calculation of the galaxy bias parameters that we show in the next section. Perhaps the only aspect of Fig.~\ref{fig:bias_halos} that may be less familiar and whose clarification is worth repeating here has to do with the appreciably different mass dependence of $b_\sigma^h$, which is monotonically decreasing, compared to that of both $b_1^h$ and $b_\phi^h$, which is monotonically increasing. As explained in Ref.~\cite{2020JCAP...02..005B}, this is due to the fact that a boost in $\Omega_b$, at fixed $\Omega_m$, slows down the growth of total matter perturbations on sub-horizon scales between radiation domination and baryon-photon decoupling. This lowers the amplitude of the total matter power spectrum after decoupling for $k > k_{\rm eq}$, which effectively suppresses the subsequent formation of nonlinear structures at later times. On the other hand, boosts in $\rho_m$ and $\mathcal{A}_s$ both enhance hierarchical  structure formation overall, hence the monotonic increase of both $b_1^h$ and $b_\phi^h$ with total halo mass. 

We have checked (not shown) that the values of $b_1^h$ and $b_\phi^h$ measured from the full Hydro versions of the simulations are consistent with those shown in Fig.~\ref{fig:bias_halos} for the Gravity case. The same is true for $b_\sigma^h$ although in this case the Hydro results tend to slightly, but systematically, underpredict the Gravity results at lower masses (cf.~Fig.~1 vs.~Fig.~2 in Ref.~\cite{2020JCAP...02..005B}). 

\section{HOD responses and their impact on galaxy bias}
\label{sec:results}

In this section we present our results on the HOD number responses and their importance in the calculation of galaxy bias using the HM. We have already exposed the problem we wish to address in the Introduction for the case of $\O = \delta_m$ perturbations, but let us re-derive Eq.~(\ref{eq:HM_bias}) again from a slightly different angle and generalize to the case of the $\O = \fnl\phi$ and $\O = \sigma$ perturbations as well. 

Within the HM framework, Eq.~(\ref{eq:ng_hm_intro}) describes the total number of galaxies $n_g$ in the Universe as the integral over the halo mass function $n_h$ weighted by the HOD number $\hodn$:
\bq\label{eq:ng_hm_results}
n_g = \int {\rm d}M_h n_h(M_h) {\hodn}.
\eq
In keeping with the response expansion of Eq.~(\ref{eq:response_def}) and the separate universe argument, the galaxy bias parameters are given by the logarithmic derivative of $n_g$ to the perturbations (or changes in cosmology) $\O$, i.e., $b_\O^g = \partial{\rm ln}\ n_g/\partial\O$. Applying this to Eq.~(\ref{eq:ng_hm_results}) gives
\bq\label{eq:bias_HM_results}
b^g_\O &=& \frac{1}{n_g}\frac{\partial}{\partial\O} \int {\rm d}M_h n_h(M_h) {\hodn} \nonumber \\
&=& \frac{1}{n_g} \int {\rm d}M_h n_h(M_h) {\hodn} \Big[\frac{\partial{\rm ln}\ n_h(M_h)}{\partial\O} + \frac{\partial{\rm ln}\hodn}{\partial\O}\Big] \nonumber \\
&=& \frac{1}{n_g} \int {\rm d}M_h n_h(M_h) {\hodn} \Big[b_\O^h(M_h) + R_\O^g(M_h)\Big],
\eq
where in the last equality we have used the definition of halo bias and the $\hodn$ responses as the logarithmic derivative of the halo abundances and HOD numbers, respectively.  For the case of $\O = \delta_m$, this equation recovers Eq.~(\ref{eq:HM_bias}). 

In our main results below, we compare the values of $b_\O^g$ measured using the separate universe simulations with the result of Eq.~(\ref{eq:bias_HM_results}), with and without the contribution from $R_\O^g(M_h)$. In our numerical calculations, we evaluate the integral of Eq.~(\ref{eq:bias_HM_results}) over the range\footnote{This is also the range of halo masses we consider in the calculation of bias parameters and responses using the separate universe simulations. We checked that for halo masses above this range our numerical results become strongly affected by the noise originating from having very few such massive objects. On the other hand, the integrand is negligible for mass scales below this range.} $M_h \in \left[10^{10}; 10^{14}\right]M_{\odot}/h$ with the quantities $n_h$, $\hodn$, $b_\O^h$ and $R_\O^g$ obtained by interpolating the simulation measurements. Our HM predictions therefore display some numerical noise inherited from the simulations. This is not ideal from a generic theory prediction perspective (for instance, one could use the solid lines in Fig.~\ref{fig:bias_halos} to describe $b_\O^h$), but it facilitates our goal to demonstrate that taking into account all of the ingredients of Eq.~(\ref{eq:bias_HM_results}) (as measured in the simulations) describes well the galaxy bias parameters (measured from the same simulations). 

In the remainder of this section, we present our numerical results in turn for (i) subhalos selected by total mass, (ii) galaxies selected by stellar mass, and (iii) for a few example gas elements tracked by the IllustrisTNG model.

\subsection{Subhalos selected by total mass}
\label{sec:results_totmass}

\begin{figure}
    \centering
    \includegraphics[width=\textwidth]{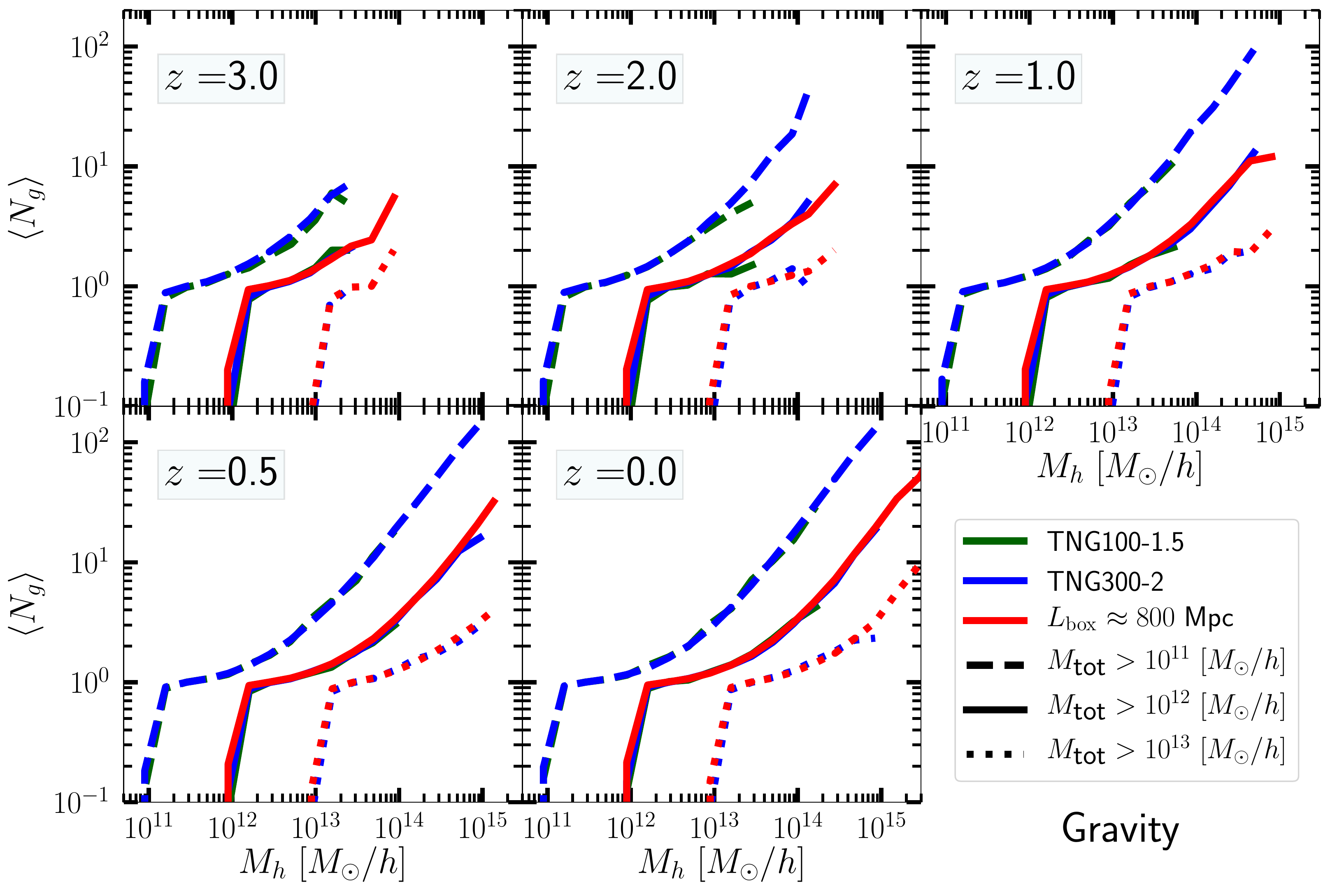}
     \caption{Mean HOD number $\hodn$ for three subhalo populations selected by their total halo mass $M_{\rm tot} > 10^{11}\ M_{\odot}/h$ (dashed), $M_{\rm tot} > 10^{12}\ M_{\odot}/h$ (solid) and $M_{\rm tot} > 10^{13}\ M_{\odot}/h$ (dotted). The different panels are for different redshifts, and the different colors are for the three numerical resolutions of the Gravity simulations, as labeled (the different resolutions agree so well that some of the lines almost completely cover the others). The $M_{\rm tot} > 10^{11}\ M_{\odot}/h$ case is not shown for the $L_{\rm box} \approx 800{\rm Mpc}$ resolution since these objects are poorly resolved. The $M_{\rm tot} > 10^{13}\ M_{\odot}/h$ is not shown for the TNG100-1.5 resolution since there are very few objects with these higher masses.}
    \label{fig:HOD_Mtot}
\end{figure}

\begin{figure}
    \centering
    \includegraphics[width=\textwidth]{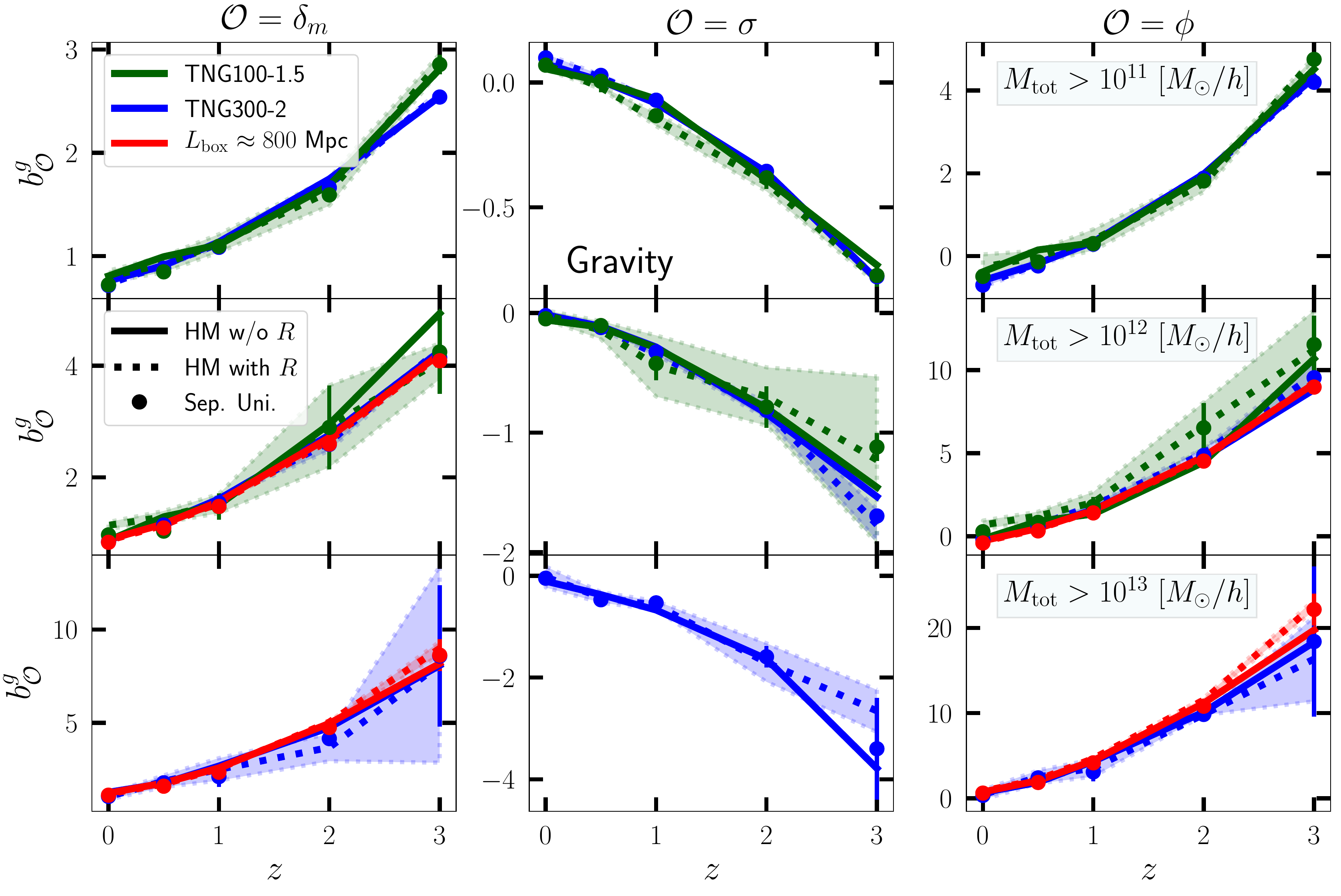}
    \caption{Galaxy bias parameters of the same total-mass-selected subhalo populations shown in Fig.~\ref{fig:HOD_Mtot} for the Gravity simulations. The left, center and right panels show the result for $b_1^g$, $b_\sigma^g$ and $b_\phi^g$, respectively. The three rows are for each of the three subhalo populations, as labeled. In each panel, the symbols with error bars show the $b_\O^g$ measured directly from the separate universe simulations, and the dashed and solid lines show the HM prediction of Eq.~(\ref{eq:bias_HM_results}) with and without the HOD number responses $R_\O^g$ taken into account.}
    \label{fig:HM_bias_Mtot}
\end{figure}

We begin our exploration of the importance of the HOD number responses $R_\O^g$ in Eq.~(\ref{eq:bias_HM_results}) by considering the case of subhalos selected by their total mass $M_{\rm tot}$ in the Gravity simulations. This is not the most realistic scenario from an observational viewpoint, but it is interesting to analyse nonetheless to subsequently compare with the results from other selection criteria below. The higher degree of numerical convergence between different gravity-only resolutions also facilitates the comparison, and in the Gravity case, we can make use also of the larger volume $L_{\rm box} \approx 800{\rm Mpc}$ set of simulations.

Figure \ref{fig:HOD_Mtot} shows the mean galaxy HOD numbers $\hodn$ measured in the simulations of the Fiducial cosmology for three subhalo total mass samples: $M_{\rm tot} > 10^{11}\ M_{\odot}/h$ (dashed), $M_{\rm tot} > 10^{12}\ M_{\odot}/h$ (solid) and $M_{\rm tot} > 10^{13}\ M_{\odot}/h$ (dotted). The result is shown for the three resolutions TNG100-1.5, TNG300-2 and $L_{\rm box} \approx 800{\rm Mpc}$ (which agree very well with one another), and at different redshifts, as labeled. The redshift and mass dependence of the HOD numbers in Fig.~\ref{fig:HOD_Mtot} is in line with the expectation. Namely, $\hodn$ transitions from zero to one at around the minimum mass of the corresponding subhalo population (i.e., the main central subhalo cannot be more massive than the host halo), and towards higher halo masses the HOD number increases in power-law fashion reflecting the larger number of satellite subhaloes that reside in massive haloes. 

The $\hodn$ curves shown in Fig.~\ref{fig:HOD_Mtot} and their response functions $R_\O^g$ (together with the halo mass function $n_h$ and halo bias $b_\O^h$) can be plugged into Eq.~(\ref{eq:bias_HM_results}) to work out the HM predictions for the galaxy (or subhalo here) bias parameters $b_\O^g$. The result is shown in Fig.~\ref{fig:HM_bias_Mtot} for $b_1^g$ (left), $b_\sigma^g$ (center)  and $b_\phi^g$ (right), and for different resolutions, redshifts and minimum mass cuts, as labeled. The symbols with error bars show the $b_\O^g$ measured directly from the separate universe simulations. The dashed lines show the outcome of Eq.~(\ref{eq:bias_HM_results}), while the solid lines show the same but with the HOD responses artificially set to zero $R_\O^g = 0$; their difference thus measures directly the importance of the $R_\O^g$. The shaded bands around the dashed lines indicate the error due to the uncertainty in the HOD response measurements alone. Overall, the figure shows that both HM calculations agree well with the separate universe results, which is indicative of a small size of $R_\O^g$ for subhalos selected by their total mass in gravity-only simulations. 

\begin{figure}
    \centering
    \includegraphics[width=\textwidth]{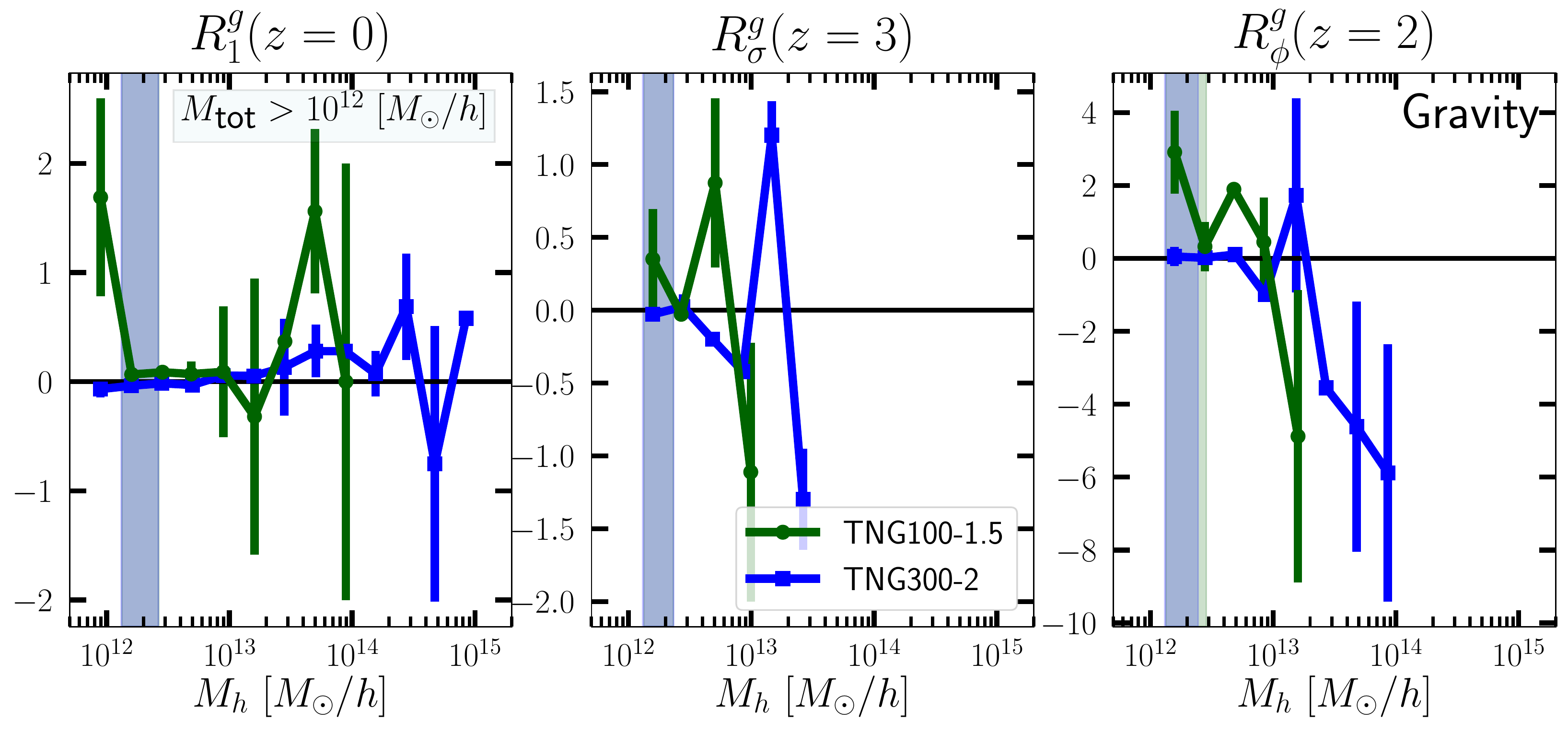}
    \caption{Example HOD number responses measured for the subhalo population with $M_{\rm tot} > 10^{12}\ M_{\odot}/h$ from the Gravity simulations. The different panels are for different perturbations $\O$ and redshifts, and the different colors for the TNG100-1.5 and TNG300-2 resolutions, as labeled. The vertical bands cover the mass range in which the weighting function of Eq.~(\ref{eq:weight}) is larger than 0.7 its maximum value; this roughly marks the range that contributes the most to the integral of Eq.~(\ref{eq:bias_HM_results}).}
    \label{fig:R_Mtot}
\end{figure}

Figure \ref{fig:R_Mtot} shows a few illustrative cases of the HOD response functions to better understand the result shown in Fig.~\ref{fig:HM_bias_Mtot} (we always limit ourselves to showing only a few illustrative examples of HOD responses for brevity). In particular, the figure shows the HOD responses measured from the TNG100-1.5 and TNG300-2 resolutions for the subhalo mass sample $M_{\rm tot} > 10^{12}\ M_{\odot}/h$ and for $\O = \delta _{m}$ at $z = 0$ (left), $\O = \sigma$ at $z=3$ (center) and $\O = \fnl\phi$ at $z = 2$ (right). The vertical shaded bands mark the range in halo mass that contributes the most to the integral of Eq.~(\ref{eq:bias_HM_results}), and therefore, where any departures of $R_\O^g$ from zero have the strongest impact. Specifically, the integrand of Eq.~(\ref{eq:bias_HM_results}) is weighted by 
\bq\label{eq:weight}
W(M_h) = n_h(M_h)\hodn,
\eq
which is negligible at both low $M_h$ (as $\hodn \to 0$) and high $M_h$ (as $n_h$ is exponentially suppressed). The width of the bands mark the range over which $W(M_h)$ is higher than 0.7 of its maximum value (note that these band widths are merely indicative and the integral is naturally also sensitive to mass scales just outside of it). Indeed, at $z=0$, the $R_1^g$ measured from both resolutions are compatible with zero near the vertical bands, hence the similarity between the dashed and solid lines at $z=0$ in the middle left panel of Fig.~\ref{fig:HM_bias_Mtot}. On the other hand, at $z=2$, the $R_\phi^g$ measured from the TNG100-1.5 resolution appears larger than zero, which explains the upwards shift at $z=2$ in the middle right panel of Fig.~\ref{fig:HM_bias_Mtot}. Similarly, at $z=3$, the $R_\sigma^g$ of the TNG100-1.5 resolution also appears larger than zero, hence the upwards shift of the dashed line at $z=3$ in the middle center panel of Fig.~\ref{fig:HM_bias_Mtot}. These small effects are however likely caused by numerical noise in the measurements of $R_\O^g$, as illustrated by the fact that the shaded bands in Fig.~\ref{fig:HM_bias_Mtot} almost always enclose the solid lines.

Overall, the results in this subsection illustrate that the HOD numbers of total-mass selected subhalos do not respond strongly to any of the three types of long-wavelength perturbations we consider in this paper. We have shown this here explicitly for the case of the Gravity simulations, but we have confirmed (not shown) that the same conclusion holds in the case of the Hydro simulations for galaxies selected by their total mass. In the subsections below, the situation becomes more interesting as the responses $R_\O^g$ become larger when the tracers are selected by properties beyond total mass.

\subsection{Galaxies selected by stellar mass}
\label{sec:results_stemass}

\begin{figure}
    \centering
    \includegraphics[width=\textwidth]{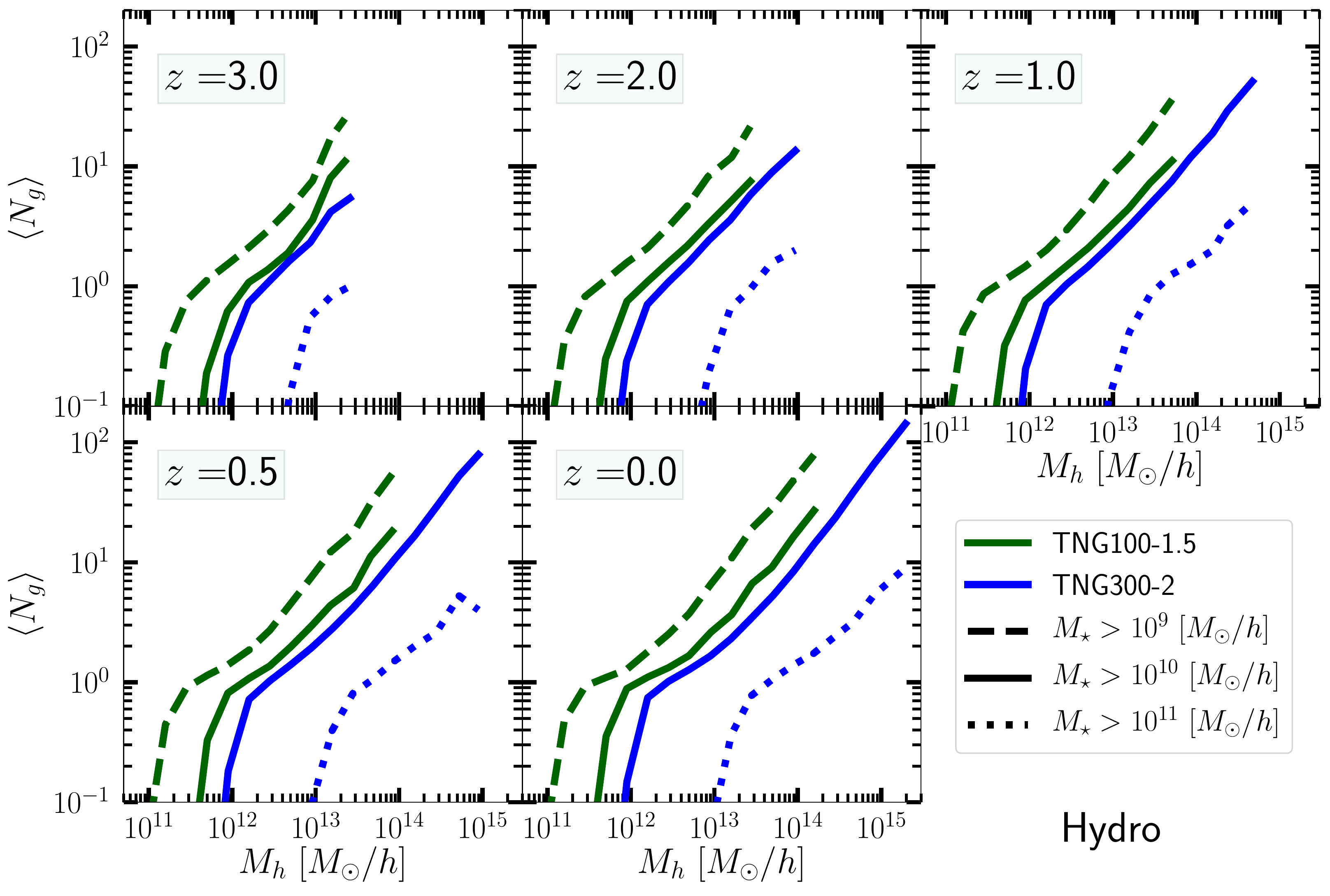}
    \caption{Mean HOD number $\hodn$ for three galaxy populations selected by their stellar mass $M_{*} > 10^{9}\ M_{\odot}/h$ (dashed), $M_{*} > 10^{10}\ M_{\odot}/h$ (solid) and $M_{*} > 10^{11}\ M_{\odot}/h$ (dotted), at different redshifts and numerical resolutions, as labeled. This is the same as Fig.~\ref{fig:HOD_Mtot}, but for stellar mass selection in the Hydro simulations, instead of total mass selection in the Gravity simulations. We do not show the $M_{*} > 10^{9}\ M_{\odot}/h$ case for the TNG300-2 resolution since this includes objects with very few star tracers and may not be well resolved. We also skip showing the $M_{*} > 10^{11}\ M_{\odot}/h$ case for the TNG100-1.5, since there are very few objects with these masses in this simulation box.}
    \label{fig:HOD_Mstar}
\end{figure}

\begin{figure}
    \centering
    \includegraphics[width=\textwidth]{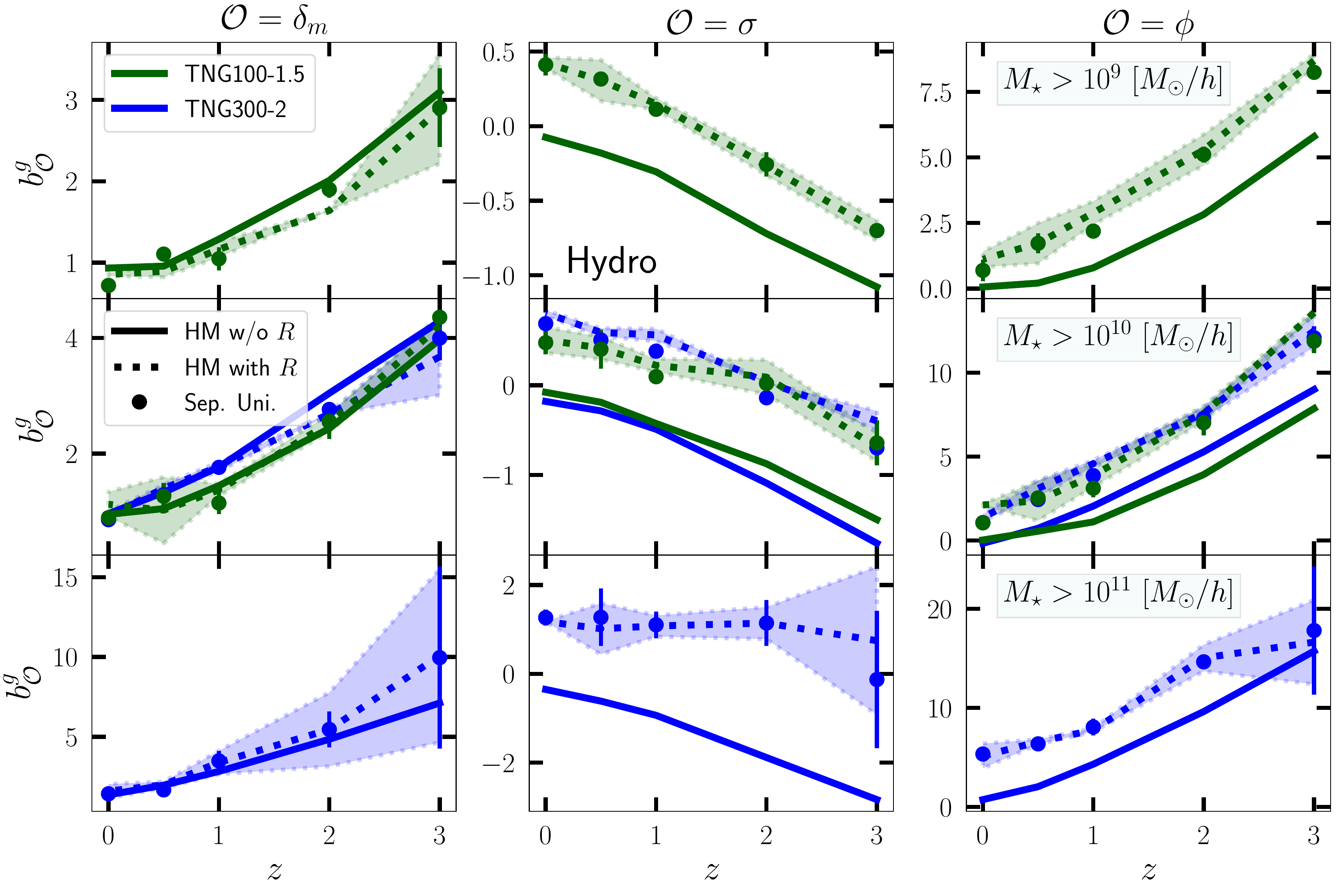}
    \caption{Galaxy bias parameters of the same stellar-mass-selected galaxy populations shown in Fig.~\ref{fig:HOD_Mstar}. This is the same as Fig.~\ref{fig:HOD_Mtot}, but for stellar mass selection in the Hydro simulations, instead of total mass selection in the Gravity simulations.}
    \label{fig:HM_bias_Mstar}
\end{figure}

We turn our attention now to the case of galaxies selected by their stellar mass. This is a more observationally relevant exercise since the total stellar mass of galaxies can be estimated from observations more robustly than their total mass. Figure \ref{fig:HOD_Mstar} shows the HOD numbers $\hodn$ for three minimum stellar mass cuts at different redshifts for the TNG100-1.5 and TNG300-2 resolutions of the Hydro simulations of the Fiducial cosmology, as labeled. The lower mass sample $M_{*} > 10^{9}\ M_{\odot}/h$ is shown only for the TNG100-1.5 resolution since these objects contain $\mathcal{O}(10)$ star particles at TNG300-2 resolution and are not therefore as well resolved. In turn, we do not show the $M_{*} > 10^{11}\ M_{\odot}/h$ case for the TNG100-1.5 resolution since the measurements are noisier as a result of having much fewer of these objects in this simulation box. 

The shape and redshift evolution of the $\hodn$ in Fig.~\ref{fig:HOD_Mstar} is in line with the expectation discussed already above in Fig.~\ref{fig:HOD_Mtot} for total mass selection. A main noteworthy difference here concerns the poorer agreement between the two resolutions: in Fig.~\ref{fig:HOD_Mtot}, the HOD numbers of the different resolutions are nearly indistinguishable, whereas in Fig.~\ref{fig:HOD_Mstar} at fixed halo mass $M_h$, the TNG100-1.5 case displays a higher number of galaxies with $M_{\star} > 10^{10}\ M_{\odot}/h$, compared to TNG300-2. This reflects the well known fact that numerical convergence is harder to achieve in hydrodynamical simulations of galaxy formation like IllustrisTNG. In particular, the higher numerical resolution of the TNG100-1.5 simulations makes star formation more efficient, which boosts the amplitude of the stellar mass function (and consequently the corresponding HOD numbers, as shown in Fig.~\ref{fig:HOD_Mstar}); see also Fig.~A.1 of Ref.~\cite{Pillepich:2017fcc} for a comparison of the stellar mass function at different numerical resolutions in IllustrisTNG. As we will see next, however, the bias and HOD response results of the two resolutions agree very well with each other. This is because the bias and HOD responses are evaluated as ratios of quantities from simulations with the same resolution, which helps to mitigate numerical resolution effects. This agreement is nontrivial and it supports the numerical convergence of our results.

\begin{figure}
    \centering
    \includegraphics[width=\textwidth]{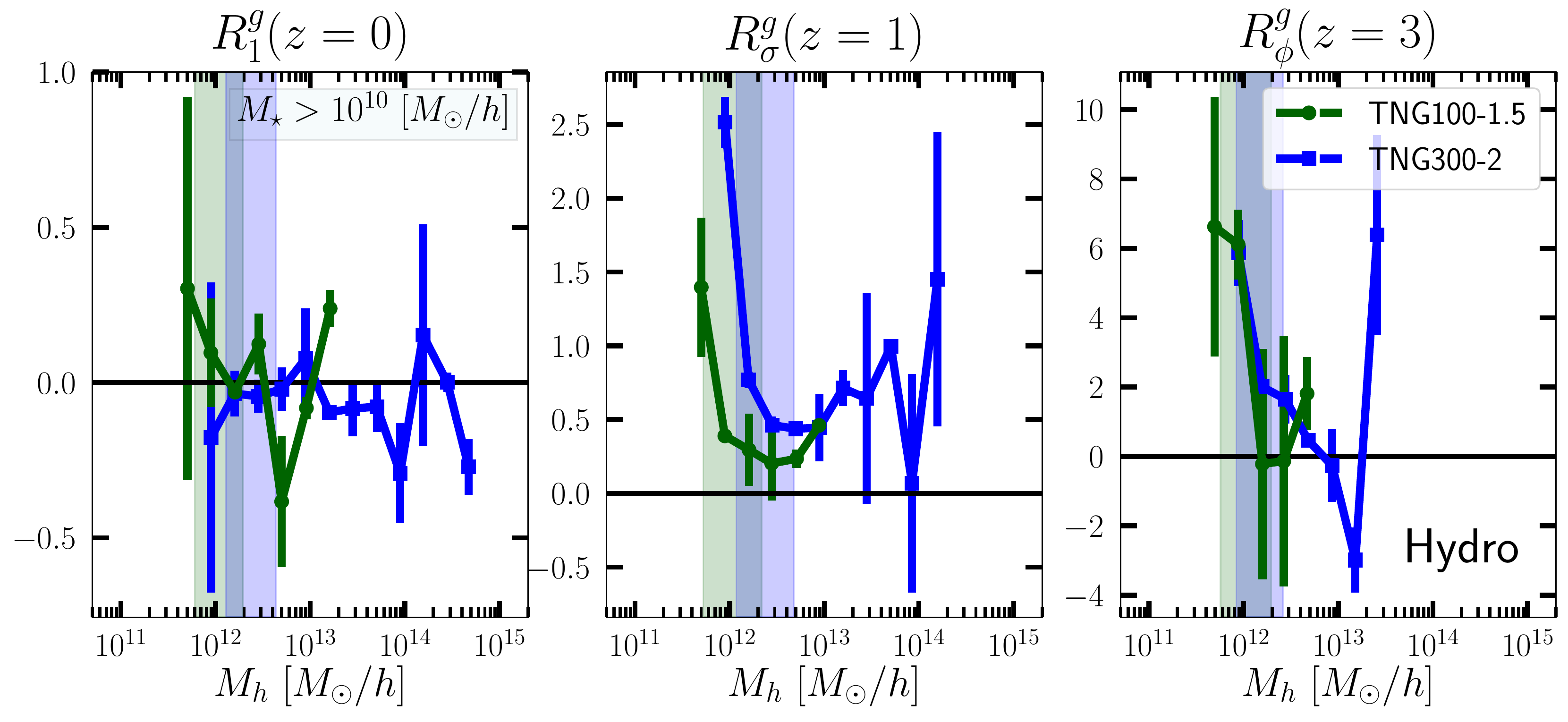}
    \caption{Example HOD number responses measured for the galaxy population with $M_{*} > 10^{10}\ M_{\odot}/h$. This is the same as Fig.~\ref{fig:R_Mtot}, but for stellar-mass selected galaxies (and different illustrative $R_\O$ at different $z$), instead of total mass selected objects.}
    \label{fig:R_Mstar}
\end{figure}

Figure \ref{fig:HM_bias_Mstar} compares the galaxy bias parameters $b_\O^g$ measured for the stellar-mass selected samples with the corresponding HM predictions of Eq.~(\ref{eq:bias_HM_results}); the latter are again shown for the cases in which the measured $R_\O^g$ are taken into account (dashed) and when they are artificially set to zero (solid). Similarly to the case of total mass selection discussed in the previous subsection, the values of $b_1^g$ under stellar mass selection do not appear dramatically affected by the HOD number response function $R_1^g$. This is seen by the comparable amplitude between the dashed and solid lines in the left panels of Fig.~\ref{fig:HM_bias_Mstar}, with both agreeing well with the separate universe simulation measurements of $b_1^g$. The left panel in Fig.~\ref{fig:R_Mstar} shows $R_1^g$ at $z=0$ for the intermediate mass sample $M_{*} > 10^{10}\ M_{\odot}/h$, and it illustrates indeed how $\hodn$ does not respond strongly to total matter density perturbations $\delta_m$ (the result is compatible with $R_1^g = 0$ within the precision of our measurements).

In contrast, the values of $b_\sigma^g$ and $b_\phi^g$ of the stellar-mass selected objects are significantly more affected by the corresponding HOD number responses, as shown by the larger difference between the dashed and solid lines in the center and right panels of Fig.~\ref{fig:HM_bias_Mstar}. The center and right panels in Fig.~\ref{fig:R_Mstar} show, as an example, $R_\sigma^g$ at $z=1$ and $R_\phi^g$ at $z = 3$, which are sizeable and vary between $[0.5-2]$ and $[1-6]$, respectively, within the halo mass range that contributes the most to Eq.~(\ref{eq:bias_HM_results}). Note that the impact of the HOD responses on $b_\sigma^g$ and $b_\phi^g$ can be quite dramatic in Fig.~\ref{fig:HM_bias_Mstar}: the values of $b_\phi^g$ become non-zero at low redshift, while the boost in $b_\sigma^g$ can be as large as to result in a change of sign at low redshift.

The reason why some $R_\O^g$ are sizeable for stellar-mass selected samples (and small for objects selected by their total mass) is interesting and can be traced back to the impact that the $\O$ perturbations have on the relation between the stellar mass and total mass of the galaxies, $M_*(M_{\rm tot})$.  Using the same simulations we use in this paper, Ref.~\cite{2020JCAP...02..005B} showed that the additional baryons supplied by CIPs provide more fuel for star formation, which works to boost the amplitude of the $M_*(M_{\rm tot})$ relation, i.e. there is more mass in stars at fixed total mass. Further, Ref.~\cite{2020arXiv200609368B}, using also the same simulations as in here, showed that the $\O = \fnl\phi$ perturbations boost the star formation rate more than they boost the total mass accretion rate, and thus, they typically enhance $M_*$ at fixed $M_{\rm tot}$, as well. On the other hand, Ref.~\cite{2020arXiv200609368B} found that $\O = \delta_m$ perturbations boost both $M_*$ and $M_{\rm tot}$ by roughly the same amount, thereby preserving the median stellar mass at fixed total mass. If $M_*(M_{\rm tot})$ is higher in the separate universe cosmology than in the Fiducial, then the same minimum stellar mass cut $M_{*, \rm min}$ in the two cosmologies corresponds to a lower minimum total mass cut $M_{\rm tot, min}$ in the separate universe cosmology. Since galaxies with lower total masses are more abundant, that explains the positive response of $R_\sigma^g$ and $R_\phi^g$ that is manifested in Figs.~\ref{fig:HM_bias_Mstar} and \ref{fig:R_Mstar}.

To make this point more specifically, recall from Eq.~(\ref{eq:bias_HM_results}) that the HOD number responses are defined as the logarithmic derivative of the HOD number w.r.t. the perturbations $\O$
\bq\label{eq:explain1}
R_\O^g(M_* > M_{*,\rm min} | M_h) = \frac{\partial{\rm ln}\langle{N}_g(M_* > M_{*,\rm min} | M_h)\rangle}{\partial\O},
\eq
where we are specializing to the case of galaxy samples selected by a minimum stellar mass cut. The corresponding HOD number can be written as 
\bq\label{eq:explain2}
\langle{N}_g(M_* > M_{*,\rm min} | M_h)\rangle &=& \int_{M_{*,\rm min}}^{\infty} {\rm d}M_* \frac{{\rm d}\langle N_g(M_{*} | M_h)\rangle}{{\rm d}M_*} \nonumber \\
&=& \int_{M_{*,\rm min}}^{\infty} {\rm d}M_* \int_0^{\infty} {\rm d}M_{\rm tot} \frac{{\rm d}\langle N_g(M_{\rm tot} | M_h)\rangle}{{\rm d}M_{\rm tot}} \mathcal{P}(M_*, M_{\rm tot}),
\eq
where ${{\rm d}\langle N_g(M_{*} | M_h)\rangle}/{{\rm d}M_*}$ denotes the number of galaxies in the stellar mass bin $\left[M_*, M_* + {\rm d}M_*\right]$ that reside in haloes with mass $M_h$, ${{\rm d}\langle N_g(M_{\rm tot} | M_h)\rangle}/{{\rm d}M_{\rm tot}}$ represents the same but for the total galaxy mass bin $\left[M_{\rm tot}, M_{\rm tot} + {\rm d}M_{\rm tot}\right]$, and $\mathcal{P}(M_*, M_{\rm tot})$ denotes the probability of a galaxy with total mass $M_{\rm tot}$ to have stellar mass $M_*$. Using Eqs.~(\ref{eq:explain2}) and (\ref{eq:explain1}), it is straightforward to demonstrate that even if the total-mass selected HOD numbers are non-responsive, i.e.,
\bq\label{eq:explain3}
\frac{\rm d}{{\rm d}\O}\left(\frac{{\rm d}\langle N_g(M_{\rm tot} | M_h)\rangle}{{\rm d}M_{\rm tot}} \right)= 0,
\eq
(which the previous subsection suggests is approximately the case) it is still possible to have non-zero responses $R_\O^g$ for the stellar-mass selected samples if the stellar-to-total-mass relation of the galaxies {\it responds}, i.e., if
\bq\label{eq:explain4}
\frac{\partial \mathcal{P}(M_*, M_{\rm tot})}{\partial\O} \neq 0.
\eq
Thus, the results from Refs.~\cite{2020arXiv200609368B, 2020JCAP...02..005B} that ${\partial \mathcal{P}(M_*, M_{\rm tot})}/{\partial\O} > 0$ for $\O = \fnl\phi$ and $\O = \sigma$, i.e., galaxies contain more mass in stars at fixed total mass, imply positive HOD number responses, which in turn causes the galaxy bias parameters to become larger, as indeed shown in Fig.~\ref{fig:HM_bias_Mstar}. Further, the finding of Ref.~\cite{2020arXiv200609368B} that ${\partial \mathcal{P}(M_*, M_{\rm tot})}/{\partial\delta_m} \approx 0$ explains why $R_1^g \approx 0$ also for stellar-mass selected galaxies. 

In Ref.~\cite{2019MNRAS.490.5693B}, using also stellar-mass selected galaxies in IllustrisTNG, the authors detect a dependence of the HOD numbers on the environment of the haloes; concretely, lower mass haloes are more likely to host a central galaxy in denser environments. Further, the authors find also that the $M_*(M_{\rm tot})$ relation is slightly boosted in denser halo environments. The apparent contradiction of these results with our finding of a negligible impact of $\O = \delta_m$ on $\hodn$ and $M_*(M_{\rm tot})$ can be explained by the fact that in Ref.~\cite{2019MNRAS.490.5693B} the environment is defined over much smaller scales than we do here. Specifically, Ref.~\cite{2019MNRAS.490.5693B} defines the environment on scales of $\approx 5\ {\rm Mpc}/h$, where the larger and nonlinear mass fluctuations can affect more strongly the number of mergers and the star formation rate, compared to the lower amplitude, linear fluctuations $\O = \delta_m \ll 1$ we consider here.\footnote{We stress also that although our results are consistent with $R_1^g \approx 0$, there is still room within the noise for the response to be nonzero. Additional realizations of separate universe simulations would be needed to beat down the statistical error and determine $R_1^g$ more precisely.}

\begin{figure}
    \centering
    \includegraphics[width=\textwidth]{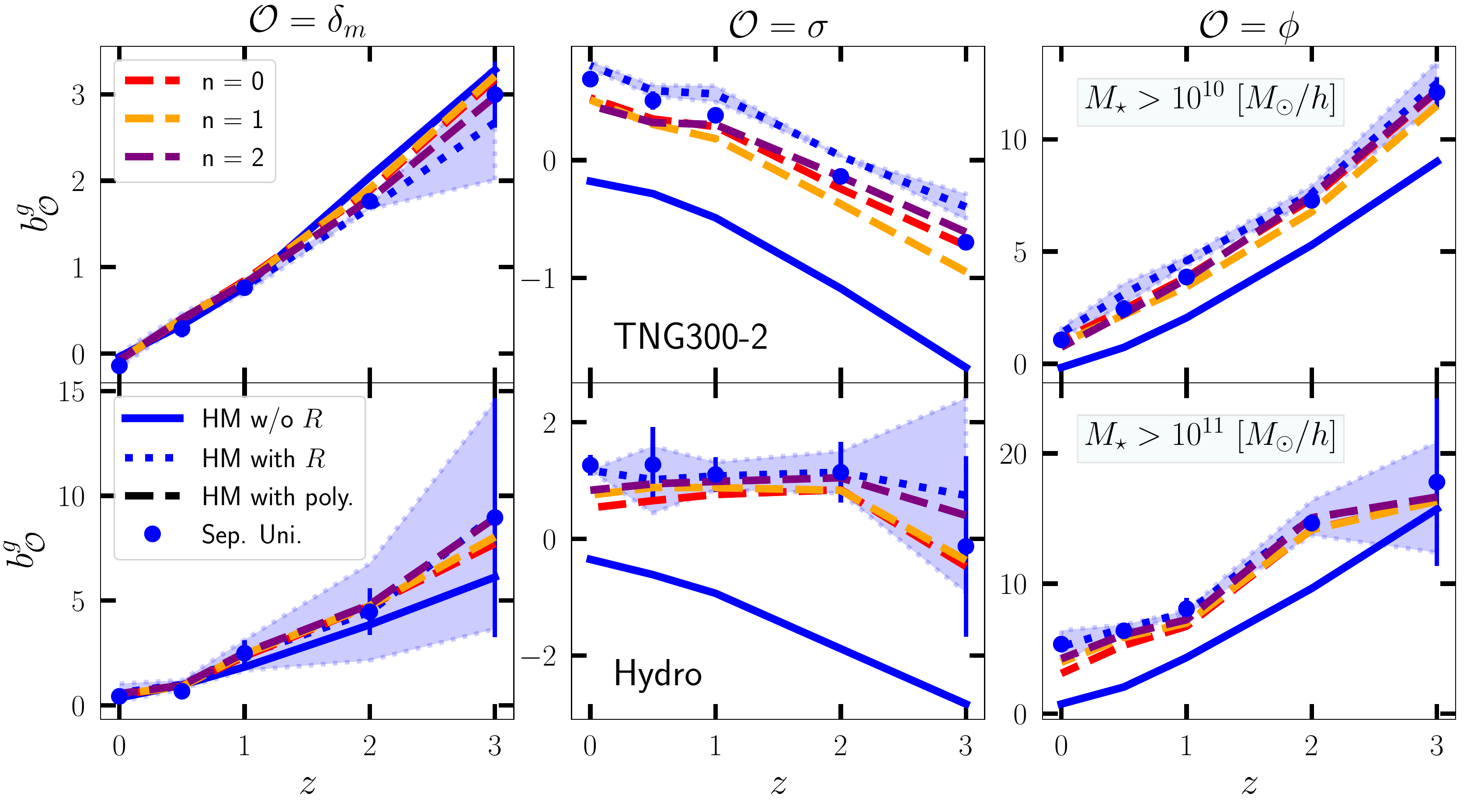}
    \caption{\small Same as Fig.~\ref{fig:HM_bias_Mstar} for the TNG300-2 resolution (the blue data points and lines are the same as in Fig.~\ref{fig:HM_bias_Mstar}), but including also the HM model prediction of Eq.~(\ref{eq:bias_HM_results}) with the measured $R_\O^{\rm g}$ replaced by polynomial fits of order $n=0, 1, 2$ (colored dashed lines), as labeled.}
    \label{fig:HM_fit_TNG300}
\end{figure}

The nonzero value of some of the HOD number responses under stellar-mass selection makes it interesting to ask how many more free parameters would be needed to describe the galaxy-halo connection in HOD studies. To roughly address this question, we fitted the $R_\O^g$ measured in the separate universe simulations with polynomials in ${\rm log}_{10}M_h$ of order $n$ over the range $M_h \in \left[10^{10}; 10^{14}\right]M_{\odot}/h$ weighted by $W(M_{h})$. These were then plugged into Eq.~(\ref{eq:bias_HM_results}) to evaluate $b_\O^g$. The results are shown in Fig.~\ref{fig:HM_fit_TNG300} for $n=0, 1, 2$ (colored dashed curves) for the TNG300-2 resolution. Interestingly, the simplest $M_h$-independent fit ($n=0$; i.e., one extra parameter describing the amplitude of $R_\O^g$) performs remarkably well for all $b_\O^g$ measurements shown, significantly improving over the $R_\O^g = 0$ case for both $\O = \fnl\phi$ and $\O = \sigma$. This is not entirely surprising since only a limited range in $M_h$ contributes strongly to the HM integral over which $R_\O^g = {\rm constant}$ has better chances of being a decent approximation. Naturally, this is just a simple exercise and more work is needed to design appropriate parametrizations of the $R_\O^g$. Nonetheless, the result in Fig.~\ref{fig:HM_fit_TNG300} does suggest that parametrizations of the $R_\O^g$ can be incorporated in HOD studies without drastically inflating the dimensionality of the parameter space of the galaxy-halo connection. This simple result adds on to efforts in the literature that aim to include environmental dependencies in HOD studies \cite{2011MNRAS.414.1207G, 2015MNRAS.454.3030P, 2016MNRAS.460.2552H, 2016arXiv160102693M, 2019MNRAS.484..989W, 2020arXiv200705545X, 2020MNRAS.491.3061S, 2020MNRAS.493.5506H, 2020arXiv200804913H, 2020MNRAS.493.5506H, 2020arXiv201004182Y}.

\subsection{Different gas elements}
\label{sec:results_gas}

\begin{figure}
    \centering
    \includegraphics[width=\textwidth]{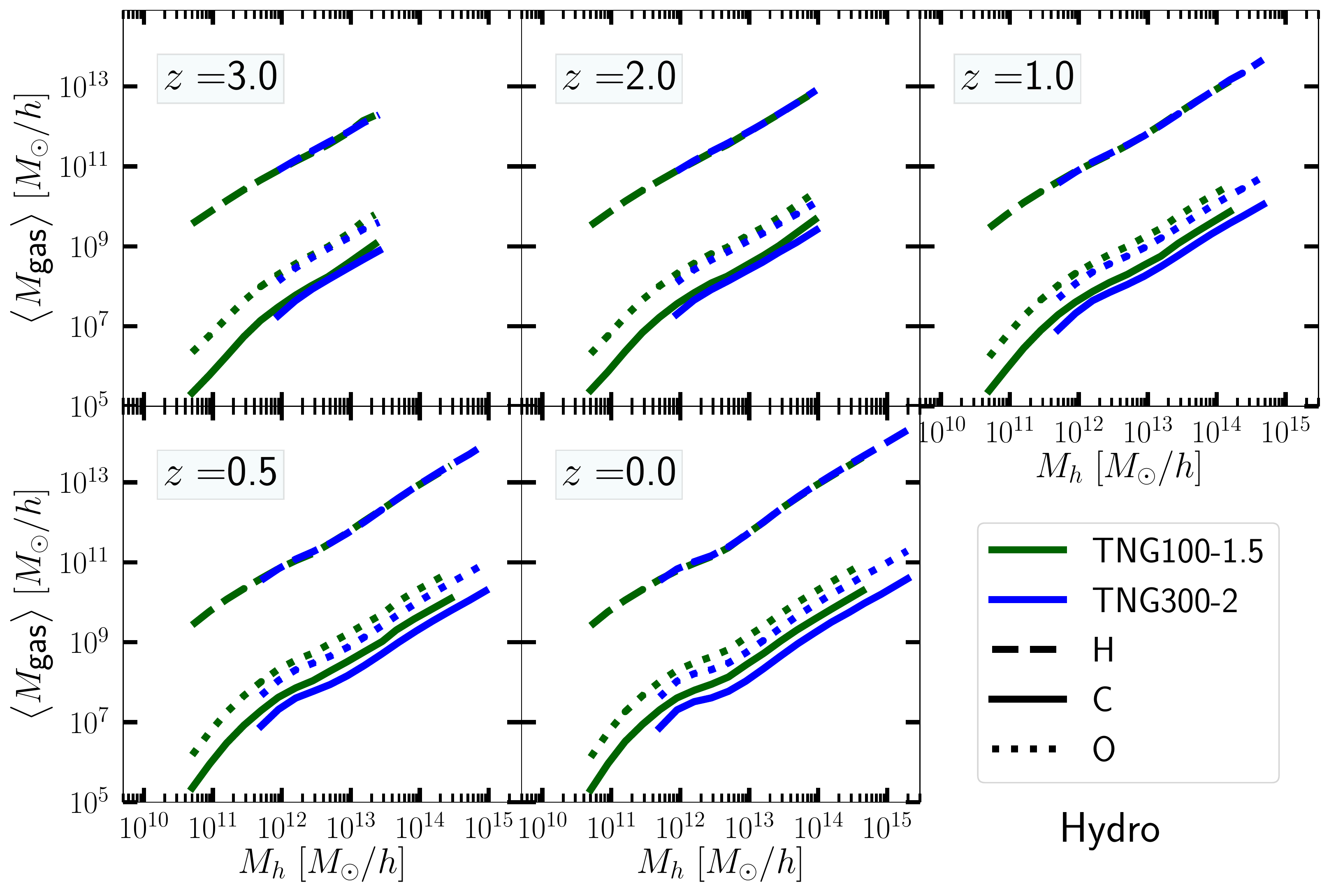}
     \caption{Mean halo gas mass as a function of total halo mass, $\langle M_{\rm gas}(M_h)\rangle$. The result is shown for hydrogen (H), carbon (C) and oxygen (O), at different redshifts and for the two resolutions TNG100-1.5 and TNG300-2, as labeled.}
    \label{fig:HOD_Gas}
\end{figure}

\begin{figure}
    \centering
    \includegraphics[width=\textwidth]{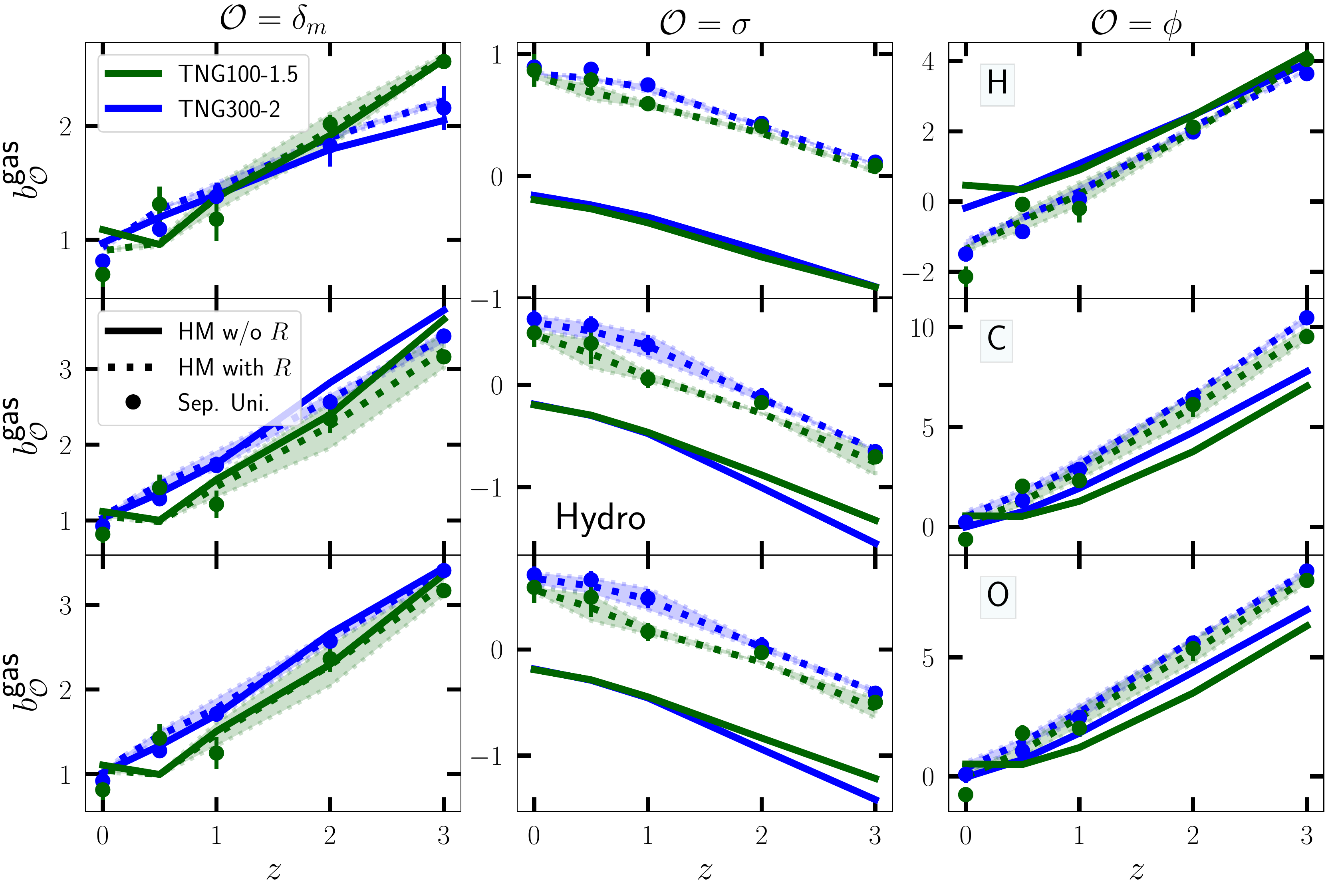}
    \caption{Bias parameters of the gas elements H, C and O shown in Fig.~\ref{fig:HOD_Gas}. This is the same as Figs.~\ref{fig:HM_bias_Mtot} and \ref{fig:HM_bias_Mstar}, but for gas as a LSS tracer, instead of galaxies selected by their total/stellar mass. }
    \label{fig:HM_bias_Gas}
\end{figure}

In this subsection, we go beyond the case of galaxies as tracers and consider the bias of the gas distribution. This is interesting observationally given the growing interest in using line-intensity mapping observations to constrain cosmology \cite{2019PhRvD.100l3522B}, which is a goal that is expected to be realized with the data from upcoming surveys such as HETDEX \cite{2008ASPC..399..115H}, CHIME \cite{2014SPIE.9145E..22B}, HIRAX \cite{2016SPIE.9906E..5XN} and SKA \cite{2020PASA...37....7S}. Rather than resolving the light emitted by individual galaxies, line-intensity mapping surveys target the integrated emission from transition lines of atoms and molecules in all galaxies and diffuse gas in the intergalactic medium. The most popular of these lines is perhaps the 21cm spin-flip transition of neutral Hydrogen \cite{2012RPPh...75h6901P, 2013ApJ...763L..20M, 2013MNRAS.434L..46S, 2015ApJ...803...21B}, but some attention is being devoted also to other emission lines from carbon monoxide (CO), ionized carbon ([CII]), oxygen (O) and Ly$\alpha$ \cite{2017MNRAS.464.1948F}.

The bias parameters of the gas $b_\O^{\rm gas}$ are defined analogously to as in Eq.~(\ref{eq:response_exp_ng}), but with the gas density $\rho_{\rm gas}$ instead of the galaxy number density $n_g$. Further, for the case of the gas distribution, instead of a mean HOD number, the relevant {\it halo occupancy quantity} is the mean mass of the gas inside halos with mass $M_h$, $\langle M_{\rm gas}(M_h)\rangle$. Its response functions are defined similarly to as in Eq.~(\ref{eq:response_exp_Ng}), and we label them as $R_\O^{\rm gas}$. The gas bias expression in the HM is given by a simple generalization of Eq.~(\ref{eq:bias_HM_results}):
\bq\label{eq:gasbias_HM}
b^{\rm gas}_\O &=& \frac{1}{\rho_{\rm gas}} \int {\rm d}M_h n_h(M_h) {\langle M_{\rm gas}(M_h, z)\rangle} \Big[b_\O^h(M_h) + R_\O^{\rm gas}(M_h)\Big].
\eq
Again, this equation is routinely used in forecast studies with $R_\O^{\rm gas}(M_h) = 0$. In this subsection we wish to determine the impact of this approximation.

We do not attempt a detailed modelling of the relevant ionized and neutral gas phases in our simulations (see e.g.~Refs.~\cite{2018ApJ...866..135V, 2018ApJS..238...33D, 2019MNRAS.482.4906P, 2020arXiv201103226S}). Rather, we consider simply a few of the gas elements as followed by the IllustrisTNG model to build intuition about $R_\O^{\rm gas}$, and leave a more detailed modelling to subsequent work. Specifically, the example gas elements we look into here are hydrogen (H), carbon (C) and oxygen (O); the corresponding $\langle M_{\rm gas}(M_h)\rangle$ are shown in Fig.~\ref{fig:HOD_Gas}. The result is in line with the expectation that $\langle M_{\rm gas}(M_h)\rangle$ is a growing function of halo mass. For the case of H (which is present at the initial conditions), this simply reflects the fact that more massive haloes accrete in general more mass (including that in the form of H). The elements C and O are released to the interstellar medium by the stellar evolution and chemical enrichment models, and so the larger mass in C and O in massive haloes follows naturally from these objects containing also more stars. This explains also why the results from the two resolutions agree well for H, but less so for C and O, whose abundance is more directly linked to star formation efficiency.

\begin{figure}
    \centering
    \includegraphics[width=\textwidth]{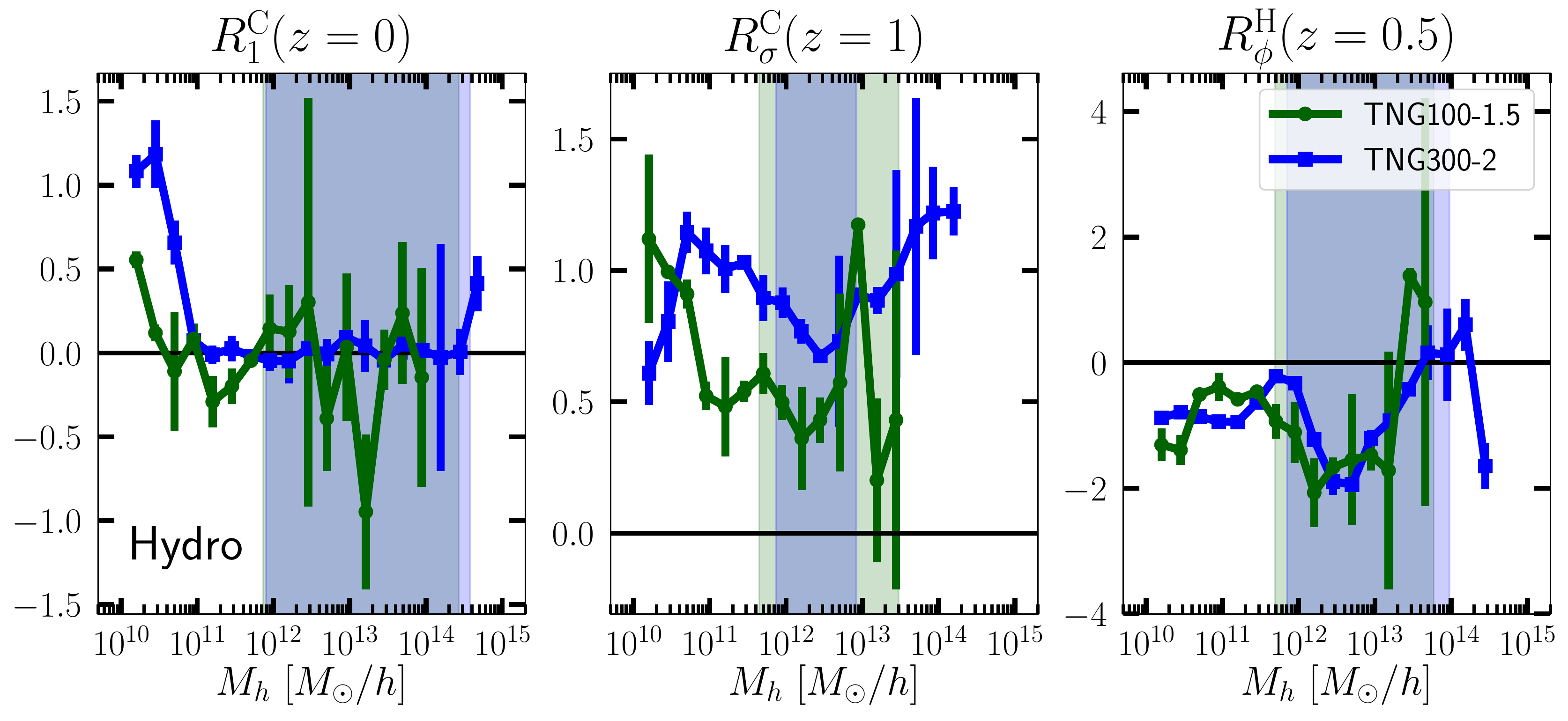}
    \caption{Example halo gas mass responses $R_\O^{\rm gas}$. This is the same as Figs.~\ref{fig:R_Mtot} and \ref{fig:R_Mstar}, but for the gas in haloes (and different illustrative $R_\O$ at different $z$), instead of galaxies selected by their total/stellar mass.}
    \label{fig:R_Gas}
\end{figure}

The gas bias parameters and a few illustrative responses $R_\O^{\rm gas}$ are shown in Figs.~\ref{fig:HM_bias_Gas} and \ref{fig:R_Gas}, respectively. The main takeaway points from these figures are similar to those from Figs.~\ref{fig:HM_bias_Mstar} and \ref{fig:R_Mstar}. In particular, the response $R_1^{\rm gas}$ has an appreciably smaller impact on the gas bias parameter $b_1^{\rm gas}$, compared to the impact of $R_\sigma^{\rm gas}, R_\phi^{\rm gas}$ on $b_\sigma^{\rm gas}, b_\phi^{\rm gas}$. The origin behind these results is also similar to that discussed in the previous subsection for stellar-mass selected galaxies. For example, for the case of the total matter perturbations $\O = \delta_m$, the boost in baryon abundance and star formation inside haloes is roughly matched by the boost in the mass of the haloes, which makes $\langle M_{\rm gas}(M_h, z)\rangle$ weakly responsive (cf.~the $R_1^{\rm C}$ result in the left panel of Fig.~\ref{fig:R_Gas}, which is compatible with zero). On the other hand, given a positive CIP $\O = \sigma$, H becomes more abundant effectively by definition of a CIP (more baryons at fixed total mass), and the additional fuel supplied for star formation also subsequently increases the abundance of C and O. This makes the halo gas mass responses $R_\sigma^{\rm gas}$ positive (the center panel of Fig.~\ref{fig:R_Gas} shows $R_\sigma^{\rm C}$ at $z=1$ as an example), which shifts the gas bias parameters $b_\sigma^{\rm gas}$ upwards in the center panels of Fig.~\ref{fig:HM_bias_Gas}.

The case of $b_\phi^{\rm gas}$ displays a couple of interesting features. For C and O, at $z>1$, the result is in line with the expectation that the increased star formation, increases the abundance of C and O, which shifts $b_\phi^{\rm gas}$ upwards. At lower redshift, however, the solid and dashed lines on the C and O panels on the right of Fig.~\ref{fig:R_Gas} approach one another, which is a consequence of a lowering of the amplitude of $R_\phi^{\rm gas}$. This can be contrasted with the $b_\phi^g$ panels of Fig.~\ref{fig:HM_bias_Mstar} for stellar-mass selected galaxies, in which the impact of the corresponding HOD response is approximately constant with redshift. Perhaps more interesting is the result for H, for which $b_\phi^{\rm H}$ is shifted downwards at lower redshift, which indicates a negative H mean mass response $R_\phi^{\rm H} < 0$ (cf.~right panel of Fig.~\ref{fig:R_Gas}). A tentative explanation here could be that the increased activity of baryonic processes such as feedback by active-galactic nuclei (AGN) could at least partially help remove some of the existing H and the excess of C and O produced by the enhanced star formation at higher redshift\footnote{Positive CIPs also increase baryonic effects' activity, but the size of the effect may be weaker compared to the $\O = \fnl\phi$ case and the larger gas fractions that are present since the initial conditions may make it harder to visualize any suppression of its abundance.}. We leave a more detailed investigation of these features to future work, in which a more robust modelling of different gas phases can also be carried out.  We anticipate that the richer the physics of the modelling of the gas phases (including the smaller-scale details of ionization), the richer the set of features that can be imprinted on $R_\O^{\rm gas}$, and consequently, on the corresponding gas bias parameters. 

Our results in this subsection demonstrate overall that the response of the gas inside haloes can be sizeable and display a few interesting nontrivial features for $\O = \sigma$ and $\O = \fnl\phi$ perturbations. We would like to emphasize that Eq.~(\ref{eq:gasbias_HM}) with $R_\phi^{\rm gas} = 0$ is routinely adopted in $\fnl$ forecast studies (see e.g.~Refs.~\cite{2019ApJ...872..126M, 2019arXiv191103964K, 2019PhRvD.100l3522B, 2020MNRAS.496.4115C, 2020arXiv201007034K} for recent examples). Our results here that $R_\phi^{\rm gas} \neq 0$ using the IllustrisTNG model therefore motivates investigations of the impact of $R_\phi^{\rm gas} \neq 0$ on $\fnl$ forecasts. Concretely, the signature from $\fnl$ on the gas power spectrum is $\propto b_1^{\rm gas} b_\phi^{\rm gas}\fnl$, and hence, $R_\phi^{\rm gas} > 0$ would drive $b_\phi^{\rm gas}$ upwards, which could potentially increase the signal-to-noise (in robust forecasts, however, uncertainties on the bias parameters should be parametrized and marginalized over \cite{2020arXiv200906622B, 2020arXiv201014523M}).

\section{Summary and Conclusions}
\label{sec:summary}

The combination of the HM with HODs is a powerful framework that is widely used to study the clustering of galaxies and construct fast galaxy mock catalogues. The HOD number $\hodn$ describes the mean number of galaxies of a given type that reside inside haloes with mass $M_h$. In this paper, we studied the dependence of $\hodn$ on the long-wavelength environment of the haloes as described by HOD response functions $R^g_\O$, where $\O$ denotes a type of long-wavelength perturbation. The $R^g_\O$ are the analog of the galaxy bias parameters $b^g_\O$, but applied to the HODs, instead of galaxy number densities; cf.~Eqs.~(\ref{eq:response_exp_ng}) and (\ref{eq:response_exp_Ng}). Despite being a natural ingredient in the HM, the HOD response functions $R^g_\O$ have remained largely ignored in the literature. This motivated us to measure them using galaxy formation simulations and study the corresponding impact in HM applications. 

We focused on the impact of the HOD responses in the HM prediction of galaxy bias $b_\O^g$. The latter is given by a weighted version of the sum of halo bias $b_\O^h$ and the HOD responses $R_\O^g$, as in Eq.~(\ref{eq:bias_HM_results}),  but in virtually all existing such calculations in the literature, the HOD responses are assumed to be zero, $R_\O^g = 0$. To test the validity of this approximation, we used separate universe simulations of the IllustrisTNG galaxy formation model to measure the HOD responses to three types of long-wavelength perturbations (cf.~Sec.~\ref{sec:sepuni}): total matter perturbations, $\O = \delta_m$; primordial potential perturbations with local PNG, $\O = \fnl\phi$; and baryon-CDM CIPs, $\O = \sigma$. Our main results consisted in comparisons between the galaxy bias parameters $b_\O^g$ measured directly from the separate universe simulations with the HM galaxy bias prediction when the measured $R_\O^g$ are appropriately taken into account or artificially set to zero. 

We have shown results from IllustrisTNG simulations run at two numerical resolutions (cf.~Table~\ref{tab:params}) and for galaxies/subhalos selected by their total mass (cf.~Sec.~\ref{sec:results_totmass}) and stellar mass (cf.~Sec.~\ref{sec:results_stemass}). Beyond galaxies as LSS tracers, we have also studied the impact of the responses of the mean gas mass in haloes $R_\O^{\rm gas}$ on the corresponding gas bias parameters $b_\O^{\rm gas}$ (cf.~Sec.~\ref{sec:results_gas}; we considered H, C and O as example case studies).

Our main conclusions can be summarized as follows:
\begin{itemize}

\item For objects selected by total mass, the HOD number responses $R_\O^g$ are compatible with zero in our simulations and impact $b_\O^g$ only weakly (cf.~Fig.~\ref{fig:HM_bias_Mtot}).

\item For stellar mass selected objects, setting $R_1^g = 0$ in the HM remains a good approximation and recovers the measured $b_1^g$ well, but doing the same for $R_\sigma^g$ and $R_\phi^g$ drastically underpredicts $b_\sigma^g$ and $b_\phi^g$ (cf.~Fig.~\ref{fig:HM_bias_Mstar}). This can be explained physically by the boost that $\O = \sigma$ and $\O = \fnl\phi$ induce on the stellar-to-total-mass relation of the galaxies (cf.~discussion in Sec.~\ref{sec:results_stemass}).

\item Likewise for the gas distribution, $b_1^{\rm gas}$ remains weakly affected by the mean mass response $R_1^{\rm gas}$, but $b_\sigma^{\rm gas}$ and $b_\phi^{\rm gas}$ are affected strongly by $R_\sigma^{\rm gas}$ and $R_\phi^{\rm gas}$ (cf.~Fig.~\ref{fig:HM_bias_Gas}). In this case, the $R_\O^{\rm gas}$ can behave differently for different gas elements (e.g., $R_\phi^{\rm C}, R_\phi^{\rm O} > 0$, but $R_\phi^{\rm H} < 0$), which suggests a nontrivial interplay of physical processes that is worth studying further. 

\item Polynomial fits to $R_\O^{g}$ of order $n \leq 2$ recover well the measured $b_\O^g$ for stellar-mass selected samples (cf.~Fig.~\ref{fig:HM_fit_TNG300}). This suggests that parametrizations of $R_\O^g$ can be added to traditional HOD studies of the galaxy-halo connection without introducing too many free parameters.  

\end{itemize}

Our results indicate overall that $R_\O^g = 0$ is an approximation that is not always valid and that it should be tested on a case-by-case basis. The sizeable effects we found on $b_\sigma^g$ and $b_\phi^g$ find immediate applications in HM-based forecast studies of CIPs and local PNG: the signals in these studies are often $\propto b_\sigma^g, b_\phi^g$, and so misestimates of the bias parameters can lead to {\it biased} conclusions. Within the precision attained by our separate universe simulation measurements, the impact of the HOD responses to total matter perturbations remained consistent with zero. As argued in Sec.~\ref{sec:results_stemass}, this is the expected result if $\delta_m$ perturbations do not modify significantly the stellar-to-total-mass relation of the galaxies. This seems to be at least approximately the case in the IllustrisTNG model (given our statistical errors), but may be different in other models of galaxy formation. It would therefore be interesting to measure $R_1^{\rm g}$ in other state-of-the-art galaxy formation models like {\sc EAGLE} \cite{2015MNRAS.446..521S, 2017arXiv170609899T}, {\sc Magneticum} \cite{2014MNRAS.442.2304H}, {\sc BAHAMAS} \cite{2017MNRAS.465.2936M}, or {\sc Horizon-AGN} \cite{2014MNRAS.444.1453D}.

We would like to emphasize the particularly welcoming aspect of the fourth bullet point above that indicates that parametrizations of $R_\O^g$ are straightforward to include in HOD studies. In Fig.~\ref{fig:HM_fit_TNG300}, we have seen that $R_\O^g = {\rm constant}$ provides a decent first order approximation, in which case only one additional parameter needs to be added to traditional HOD studies. It is important to note also that in this paper we have assumed that the HODs and their responses depend only on halo mass $M_h$, but it would be interesting to study $R_\O^g$ as a function of additional properties such as halo concentration, halo formation time, halo spin, etc.

Our analysis can be extended further to include a selection of tracers that resembles more closely the observations, including for example, an improved modelling of the gas phases that are relevant for future line-intensity mapping studies, or galaxies selected by properties such as star formation rate, color or AGN luminosity. With additional separate universe simulations, it would be also possible to study higher-order response functions (which are defined as higher-order derivatives w.r.t.~$\O$) and/or response functions associated with additional perturbations such as large-scale tidal fields $\O = K_{ij}(\vx)K^{ij}(\vx)$ \cite{mcdonald/roy:2009, chan/scoccimarro/sheth:2012, baldauf/etal:2012, saito/etal:14, andreas, 2020arXiv200306427S, 2020MNRAS.496..483M, 2020arXiv201106584A}, higher-derivative operators $\O = \nabla^2\delta_m(\vx)$ \cite{2019JCAP...11..041L} or the ionizing radiation field \cite{2017PhRvD..96h3533S, 2019JCAP...05..031C}.

\acknowledgments
We would like to thank Sownak Bose, Giovanni Cabass, Elisabeth Krause, Dylan Nelson, Annalisa Pillepich and Fabian Schmidt for useful comments and discussions at various stages of this work. RV is supported by FAPESP. AB acknowledges support from the Starting Grant (ERC-2015-STG 678652) “GrInflaGal” from the European Research Council, and from the Excellence Cluster ORIGINS which is funded by the Deutsche Forschungsgemeinschaft (DFG, German Research Foundation) under Germany's Excellence Strategy - EXC-2094-390783311. The numerical analysis of the simulation data presented in this work was done on the Cobra supercomputer at the Max Planck Computing and Data Facility (MPCDF) in Garching near Munich.

\bibliographystyle{utphys}
\bibliography{REFS}

\end{document}